\documentclass[%
 reprint,
 amsmath,amssymb,
 aps,onecolumn,
]{revtex4-2}

\usepackage{graphicx}
\usepackage{dcolumn}
\usepackage{bm}
\usepackage{tikz}
\usetikzlibrary{calc}
\usetikzlibrary{shapes}

\begin{document}

\preprint{APS/123-QED}

\title{Direct numerical simulations of a cylinder cutting a vortex}

\author{Steven Soriano}
\affiliation{NASA Glenn Research Center, Cleveland, OH 44135, USA}
\affiliation{Cullen College of Engineering, University of Houston, Houston, TX 77204, USA}
 
\author{Rodolfo Ostilla-M\'onico}
\email{rodolfo.ostilla@uca.es}
\affiliation{Dpto.~Ing. Mec\'anica y Dise\~no Industrial, Escuela Superior de Ingenier\'ia, Universidad de C\'adiz, Av.~de la Universidad de C\'adiz 10, 11519 Puerto Real, Espa\~na}
\affiliation{Cullen College of Engineering, University of Houston, Houston, TX 77204, USA}

\date{\today}

\begin{abstract}
The interaction between a vortex and an impacting body which is oriented normally to it is complex due to the interaction of inviscid and viscous three-dimensional mechanisms. To model this process, direct numerical simulations of a thin cylinder intersecting a columnar vortex oriented normally to it are conducted. By varying the impact parameter and the Reynolds number, the two regimes of interaction mentioned in the literature are distinguished: the weak and strong vortex regimes. Low impact parameters, representing strong vortices, led to ejection and interaction of secondary vorticity from the cylinder's boundary layer, while high impact parameters, representing weak vortices, led to approximately inviscid interaction of the cylinder with the primary vortex through deformations. No significant effect of the Reynolds number in the overall phenomenology is found, even if larger Reynolds numbers lead to the formation of increasingly smaller and more intense vortex structures in the parameter range studied. Finally, the hydrodynamic force curves on the cylinder are analyzed, showing that intense forces could be locally generated for some parameter regimes, but that the average force on the cylinder does not substantially deviate from baseline cases where no vortex was present. Our results shed light on the underlying mechanisms of vortex-body interactions and their dependence on various parameters.
\end{abstract}

\maketitle

\section{Introduction}
\label{chap:intro}

Vortices are one of the most characteristic structures present in fluid flows, and they have been referred to as ``the sinews and muscles of fluid dynamics'' \cite{saffman1995vortex}. At high Reynolds numbers, vortices are generally modelled as singular lines or sheets. A wide variety of flows such as aircraft or propeller-tip wakes or bubble rings can then be studied through the composition of singular solutions \cite{saffman1995vortex}. Much attention has been given to how vortex tubes, rings or sheets interact with each other or with solid boundaries, including changes in the topology of singular solutions such as those seen in reconnecting vortex rings \cite{kida1994vortex}. However, the purposeful breaking or ``cutting'' of topological vortex lines with solid objects has seen less work due to experimental and numerical challenges \cite{marshall1994vortex,naitoh1995vortex,liu2004blade}. This flow phenomenology finds significant relevance in a specific context: blade-vortex interaction (BVI), which involves the intricate physical interaction when a rotating blade encounters a nearby vortex tube \cite{rockwell1998vortex}, but is also present in other physical systems such as Type-II superconductors \cite{palau2008vortex,glatz2016vortex}. 

BVI holds widespread significance in various engineering applications, spanning wind turbines, helicopter rotors, marine propellers, and industrial fans \cite{rockwell1998vortex}, and due to its practical applications has been the workhorse of research in vortex line cutting in fluid dynamics. The majority of research on BVI concentrates on aerodynamic blades intersecting a straight vortex tube. This specific BVI scenario, referred to as perpendicular or normal BVI, occurs when the vortex and blade axes are perpendicular, and the blade cleaves through the vortex tube. Normal BVI is commonly encountered in applications such as helicopter rotors and wind turbines, where rotating blades confront vortices originating from prior blades or generated by their own motion \cite{jameson1970analysis,witkowski1989aerodynamic, leishman1998challenges, yung2000rotor,amet20092d}. Other orientations of the blade in relation to the vortex are possible, resulting in other types of BVI: grazing BVI, where the blade is perpendicular to the tube, but it moves in the direction of the core, and parallel BVI \cite{wilder1998parallel}, where the blade is parallel to the tube, and cuts the tube throughout its extent rather than at a single point. These are outside the scope of our study which focuses on swift cutting of the core.

Experiments and simulations have shown that the interaction between a vortex tube and a body such as a blade or a cylinder depends on various parameters, such as body velocity, vortex core radius, vortex swirl velocity, body geometry and size, and turbulence within the body's boundary layer \cite{Ahmadi1986AnEI,  kim1995summary, krishnamoorthy1998three, gossler2001simulation, felli2009experimental, peng2015vortex, felli2021underlying}. In particular, for a body impacting a vortex with no axial flow, three dimensionless parameters are relevant. These parameters are: (1) the impact parameter ($I_P=2\pi\sigma_0 V/\Gamma$), which is the ratio of the body's free-stream velocity to the maximum vortex swirl velocity, where $\sigma_0$ is the radius of the vortex, $V$ the velocity of the body, and $\Gamma$ the vortex's circulation; (2) the vortex Reynolds number ($Re_\Gamma=\Gamma/\nu$), where $\nu$ is the kinematic viscosity; and (3) the body thickness parameter ($T$), which is the ratio of a characteristic length such as the blade's curvature or the cylinder diameter $D$, to the vortex core radius $\sigma_0$ \cite{marshall1997instantaneous, krishnamoorthy1998three}.

Depending on both the response of the primary vortex to the body and about the interaction of the vortex with the body's boundary layer, response regimes can be constructed depending on the value the dimensionless parameters take. The response regimes are commonly referred to as the weak vortex regime, the strong vortex regime, and the bending regime. In this manuscript, we will simulate cases in the weak and strong vortex regime, while the bending regime will not be investigated. However, we discuss here all three regimes for completeness.

The weak vortex regime occurs for thin bodies $(T=D/\sigma_0 \lesssim 1)$ when the vortex is sufficiently weak $(I_P > 0.2)$. This regime is characterized with very minimal boundary layer separation and ejection until after the body has penetrated the primary vortex \cite{marshall1997instantaneous}. Once the boundary layer has separated, the separated vorticity becomes entrained within the primary vortex and does not wrap around it, instead spreading away from the body and into the vortex core. 

The strong vortex regime has been seen to occur when the vortex is sufficiently strong $(I_P<0.08)$ and is independent of the thickness parameter \cite{krishnamoorthy1998three}.  This regime is characterized with a large amount of interaction between the body's boundary layer and the vortex prior to the body's leading edge making contact with the vortex. The ejected vorticity from the body's boundary layer will roll up into a series of vortex loops that will wrap around the primary vortex. The wrapping of the secondary vorticities creates a series of wave motions that will eventually disrupt the primary vortex. This procedure of primary vortex disruption may also occur when the body is several core radii away from the vortex, but only when the value of the impact parameter is relatively low \cite{marshall1994vortex}.    

The bending regime occurs when the vortex is sufficiently weak $(I_P>0.2)$ and the body is sufficiently thick $(T>5)$ \cite{affes1993model_part_1,marshall1997instantaneous}. Separation of the boundary layer does not occur before the body has travelled through the original position of the vortex. This regime also sees large scale deformation of the primary vortex due to the inviscid interaction with the body. The core radius of the vortex will thin near the region where the body will impact the vortex due to stretching of the flow about the body. Once the vortex has deformed around the body, the body's boundary layer will separate and generate wave motions within the vortex eventually leading to a disruption of the vortex \cite{marshall1994vortex, krishnamoorthy1998three}. 

While experimental studies of this process exist, experiments are limited in their access to the full velocity and pressure fields. Furthermore, the effect of turbulence in the boundary layer on the interaction process is not well understood, partly due to the challenges in experimentally controlling inlet turbulence. Numerical simulations could in principle allow a detailed study of the development and evolution of the vortex formation and boundary layer detachment, determining the spatial and temporal forces present on the blade, as well as having full control over the turbulence in the system. They also have the advantage of being able to controlling input parameters separately, i.e.~varying $Re_\Gamma$ separately from the impact parameter, which is not easy in an experiment.

Inviscid filament models are the oldest method used to simulate BVI, due to their straightforwardness and effectiveness but present well-known limitations: they require extensions to account for variations in the vortex core area and to ensure no-penetration conditions on the surfaces; and they are inviscid, which means that they cannot be used to analyze vortex-induced boundary layer separation \cite{moore1972motion, lundgren1989area,marshall1991general,affes1993model_part_1,marshall1994vortex}, a crucial phenomena in the weak vortex regime. Nonetheless, they are useful for predicting how the body will penetrate the vortex or how the vortex will react to the motion of the body, and can also been used to estimate the pressure on the body surface during the interaction with a vortex up to a certain distance from the body \cite{affes1993model_part_1,affes1993model_part_2}. A number of extensions to the filament model which incorporate two-dimensional boundary layer dynamics have been conducted, which hint at the type of complex flow phenomena which arise \cite{doligalski1984boundary,Luton1995InteractionOS}. These were later extended to include three-dimensional boundary layers \cite{van1990lagrangian,affes1994boundary}. However, they remain limited by the fact that the underlying filament models present singularities and cannot fully resolve the dynamics \cite{affes1994boundary}.

The highest fidelity simulations of normal BVI available in the literature employ three-dimensional direct numerical simulations of a vortex impacting a blade \cite{liu2004blade, saunders2015vorticity}. These studies have detailed the way a vortex cuts through a blade, and the way this cutting generates circulation and a lift force. Ref.~\cite{liu2004blade} was limited in the range of Reynolds numbers they could simulate, as well as the choice of large impact parameters due to the lack of adequate computational power. Ref.~\cite{saunders2015vorticity} simulates larger Reynolds numbers by fixing the blade Reynolds number to be $Re_b=1000$, and varying the strength of the vortex. This study focuses on large impact parameters, $I_P>0.5$, and does not access the strong vortex regime.

As can be gathered from above, a detailed numerical study of the separation process in the weak vortex regime and the transition from weak to strong vortex regimes is missing. The use of DNS in vortex-body interaction studies has the potential to clarify previously unclear areas such as the viscous interaction of the body with the vortex \cite{saunders2015vorticity}. Simulations also allow careful control of the Reynolds number and can assess questions of Reynolds number independence unanswered in experiments \cite{krishnamoorthy1998three}.
In this manuscript, we set to address these knowledge gaps by conducting three-dimensional direct numerical simulations (DNS) of tube cutting using state-of-the-art numerical techniques. We will simulate a simplified problem: a thin wire (cylinder) impacting normally a high-Reynolds number vortex. We will study the interaction process by varying two parameters: the impact parameter $I_P$ and the circulation Reynolds number $Re_\Gamma$. Values of $I_P$ will be chosen such that simulations can access the weak and strong vortex regimes, while varying $Re_\Gamma$ will modify the secondary vorticity of the cylinder, as well potentially triggering more hydrodynamical instabilities in the primary tubes which set in at values of $Re_\Gamma\sim\mathcal{O}(10^3)$ \cite{leweke2016dynamics}. We will focus on key flow phenomena observed during the interaction, calculate the induced force on the cylinder as it cuts through the vortex, and relate how major flow structures influence the force curve. 

The manuscript is organized as follows: Section \ref{sec:num} details the numerical methods used in the manuscript. Section \ref{sec:euler} presents the main results from the flow field, including the transition between strong and weak vortex regimes (\ref{sec:transition}), and in-depth analysis of the strong (\ref{sec:strongv}) and weak vortex regimes (\ref{sec:weakv}). Section \ref{sec:force} examines the force in the wire. Finally, Section \ref{sec:conc} presents a summary of the findings as well as an outlook for future work.

\section{Numerical methods}
\label{sec:num}

\begin{figure}[ht]
\begin{tikzpicture}
\pgfmathsetmacro{\x}{-4}
\pgfmathsetmacro{\y}{-4}
\pgfmathsetmacro{\z}{-4}
\path (0,0,\y) coordinate (A) (\x,0,\y) coordinate (B) (\x,0,0) coordinate (C) (0,0,0)
coordinate (D) (0,\z,\y) coordinate (E) (\x,\z,\y) coordinate (F) (\x,\z,0) coordinate (G)
(0,\z,0) coordinate (H);

\filldraw[color=black, fill=gray!60] (-3.5,-2,-4) circle (0.15);
\fill[fill=gray!60] (-3.606,-1.893,0) -- (-3.606,-1.893,-4) -- (-3.393,-2.106,-4) -- (-3.393,-2.106,0) -- cycle;
 \draw[black] (-3.606,-1.893,0) -- (-3.606,-1.893,-4);
 \draw[black] (-3.393,-2.106,0) -- (-3.393,-2.106,-4);
\filldraw[color=black, fill=gray!60] (-3.5,-2,0) circle (0.15);

\draw[line width=0.05mm,dash dot,color=black] (-3.65,-2,0) -- (-3.65,-1.4,0) ;
\draw[line width=0.05mm,dash dot,color=black] (-3.35,-2,0) -- (-3.35,-1.4,0) ;
\draw[line width=0.05mm,<->, color=black] (-3.35,-1.5,0) -- (-3.65,-1.5,0);
\node[fill=none] at (-3.5,-1.25,0) {$D$};

\draw [black,-stealth] (-3,-2,-2) -- (-2.5,-2.0,-2) node [right] {$V$};

\draw [-latex, black, rotate=90] (-1.5,-0.9) arc [start angle=-120, end angle=160, x radius=0.25, y radius=0.5] node [below left] {$\Gamma$};

\fill [left color=gray!10,right color=gray!10,middle color=gray!60,shading angle=90] (0.4,-3.25) rectangle (0.6,0.75);
\draw [black, dash dot] (0.5,-3.25) -- (0.5,0.75);

\draw[thick,->] (-4.5,-3.5) -- (-4,-3.5) node[below]{$x$};
\draw[thick,->] (-4.5,-3.5) -- (-4.5,-3.) node[left]{$y$};
\draw[thick,->] (-4.5,-3.5) -- (-4.15,-3.15) node[above]{$z$};

\end{tikzpicture}
\caption{Schematic view of the computational domain. The $z$ direction is shortened for clarity.}
\label{fig:comp_domain}
\end{figure}
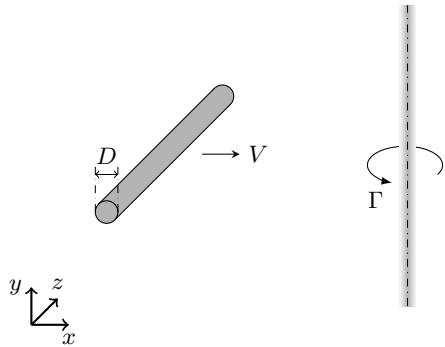

To study BVI, we directly simulate the the incompressible Navier-Stokes equations:

\begin{equation}
     \displaystyle\frac{\partial\textbf{u}}{\partial t} + \textbf{u}\cdot\nabla\textbf{u} = -\rho^{-1}\nabla p +\nu\nabla^2 \textbf{u}+ \textbf{f} 
 \label{eq:ns}
\end{equation}

\begin{equation}
 \nabla\cdot \textbf{u}=0
 \label{eq:compress}
\end{equation}

\noindent where $\textbf{u}$ is the velocity, $t$ is time, $\rho$ is the fluid density, $\nu$ is the fluid kinematic viscosity and $\textbf{f}$ is a body force originating from the the immersed boundary method (IBM) forcing used to model the cylinder. A schematic of the system is presented in Figure \ref{fig:comp_domain}, which shows a cylinder of diameter $D$, with an axis parallel to the $z$-axis travels in the $x$ direction at a velocity $V$ towards a vortex of core-size $\sigma_0$ which has an axis parallel to the $y$ direction. This vortex is initialized at the initial time with a Lamb-Ossen (Gaussian) velocity profile with circulation $\Gamma$. The vortex circulation $\Gamma$ and the vortex radius $\sigma_0$ are used to non-dimensionalize the system. 

The computational domain is taken as triply periodic, with periodic lengths $\mathcal{L}_x \times \mathcal{L}_y \times \mathcal{L}_z= 30\sigma_0 \times 30\sigma_0 \times 90\sigma_0$. The larger length in $\mathcal{L}_z$ allows for the force on the object to return to baseline values, this effect is fully discussed in Section \ref{sec:force}. The simulation is initialized with a zero velocity field except for the Lamb-Ossen vortex with the core extending in the $y$-direction and centered at ($x=0$, $z=0$). The cylinder starts from rest, and travels for $10\sigma_0$ at $V$ velocity before reaching the vortex's core line. Because the distance is the same, this means that for varying cylinder velocity (through changing $I_P$), the cylinder reaches the vortex at different times. After cutting the axis of the vortex, the cylinder then travels to the opposing end of the box. This allows us to capture all stages of the interaction with the vortex. 

The diameter of the cylinder is fixed as $0.95\sigma_0$, keeping the thickness parameter $T=D/\sigma_0\approx 1$ in the thin regime. The impact parameter is varied between $I_P\in(0.05,0.25)$. With this value of $T$, the degree of (inviscid) vortex bending that occurs before the ejection of boundary layer vorticity is determined by the impact parameter. The range of values of $I_P$
chosen, which allows us to explore the weak- and strong interaction regimes for low and high values of $I_P$ respectively. Finally, the circulation-based Reynolds numbers are varied in the range $Re_\Gamma \in(1000,3000)$. This will impact the flow in two ways: the boundary layer of the cylinder will be thinner with higher vorticity, and the increased Reynolds number may also trigger hydrodynamical instabilities in the primary tube, as many instabilities in tubes set in at values of $Re_\Gamma$ which are between $1000$ and $3000$ \cite{leweke2016dynamics}. 

The associated cylinder Reynolds number $Re_c = V D/\nu$ is calculated as $Re_c=\Gamma  I_P  T / (2\pi \nu)=Re_\Gamma I_P T/(2\pi)$, and varies in the range $Re_c\in(7.96,119)$. Table \ref{tab:sum} summarizes the control parameters used for all the runs in this manuscript. We note that while the values of $Re_c$ used here appear substantially lower than those used in Ref.~\cite{saunders2015vorticity}, the reference uses a chord-based Reynolds number, while this manuscript uses something analogous to a ``thickness''-based Reynolds number leading to an order of magnitude discrepancy when comparing $Re_c$ to the blade Reynolds number in Ref.~\cite{saunders2015vorticity}. 

\begin{table}
    \centering
    \begin{tabular}{|c|c|c|c|}
    \hline
    $T$ & $I_P$ & $Re_\Gamma$ & $Re_c$ \\ \hline
    0.95 & 0.05 & 1000 & 7.56 \\
    0.95 & 0.15 & 1000 & 22.7 \\
    0.95 & 0.25 & 1000 & 37.8 \\
    0.95 & 0.05 & 2000 & 15.1 \\
    0.95 & 0.25 & 2000 & 75.6 \\
    0.95 & 0.05 & 3000 & 22.7 \\
    0.95 & 0.25 & 3000 & 113 \\
    \hline 
    \end{tabular}
    \caption{Control parameters for the simulations of tube-wire interaction in this manuscript.}
    \label{tab:sum}
\end{table}

Simulations are ran using the open-source code AFiD \cite{van2015pencil}. This code uses energy-conserving second order centered finite differences to spatially discretize the domain. Time marching is done through a third-order Runge-Kutta for the non-linear terms and a second order Crank-Nicolson for the viscous terms. The spatial resolution is taken as $624\times624\times1872$, meaning the grid is equally spaced in all directions, with a grid length $\sim 0.05\sigma_0 \approx 0.05D$. Resolution adequacy is checked by monitoring numerical dissipation to be less than $2\%$. To increase the confidence in the mesh adequacy, Appendix \ref{app:grid} shows a comparison between a coarser mesh, the mesh used in this manuscript and a finer mesh for the $I_P=0.25$, $Re_\Gamma=3000$ case.

The cylinder is incorporated through the immersed boundary method (IBM) using a moving-least-squares formulation \cite{spandan2017parallel}. This algorithm has been used previously for turbulent simulations with rigid, moving and deforming boundaries, and has been well validated \cite{spandan2018physical,zhu2018wall,berghout2019direct}. The surface is discretized using $388800$ triangles. The triangle skewness is $<0.3$, and the average edge length is about $70\%$ of the grid spacing. This ensures adequate resolution without consuming excessive computational resources \cite{spandan2017parallel}.

\section{Flow Dynamics for Tube-Wire interaction}
\label{sec:euler}

\subsection{The distinction between weak and strong vortex regimes}
\label{sec:transition}

This subsection examines cases at $Re_\Gamma=1000$, varying $I_P$ to illustrate the difference between the strong and weak vortex regimes. The $Q$-criterium \cite{hunt1988eddies,meneveau2011lagrangian} will be used to highlight the primary vortex. This is defined as:

\begin{equation}
 Q=-\frac{\partial u_j}{\partial x_i} \frac{\partial u_i}{\partial x_j} = \frac{1}{2}\omega^2 - \sigma^2
 \label{eq:Q}
\end{equation}

\noindent where $\omega^2=(\nabla \times \textbf{u})^2$ is the vorticity magnitude squared, and $\sigma_{ij}=\frac{1}{2}(\partial_i u_j + \partial_j u_i)$ the strain tensor, with $\sigma^2=\sigma_{ij}\sigma_{ij}$. $Q$ represents the source term in the pressure-Poisson equation \cite{pumir1994numerical} and is useful for distinguishing rotation-dominated regions of the flow, which will tend to have positive values of $Q$, from strain-dominated regions which have negative values of $Q$. Visualizing regions of positive $Q$ through time will highlight the dynamics of the primary vortex, which is rotation-dominated, while leaving out the boundary layers. Later, $|\omega|$ visualizations will be presented to emphasize the dynamics of the wire's boundary layer and secondary vortices and its interaction with the primary vortex.

\begin{figure}
    \centering
    \includegraphics[width=\textwidth]{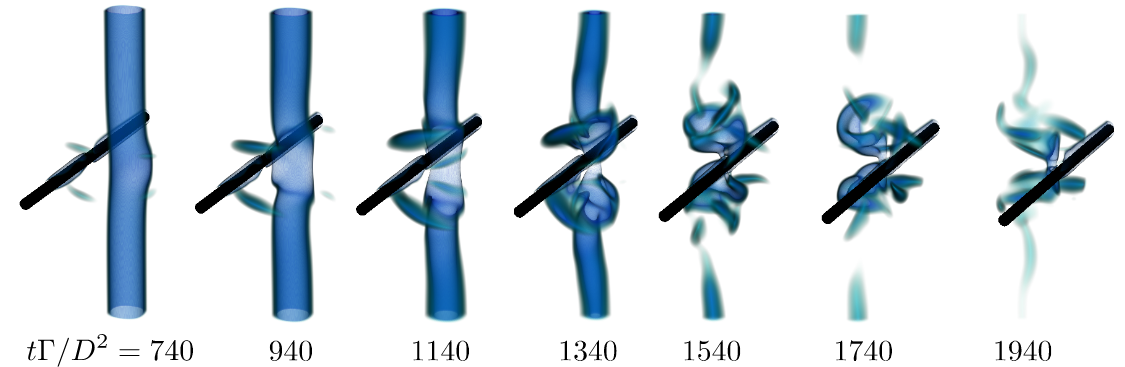}
    \caption{Volume rendering visualization of the time evolution of $Q$ (Eq.~\ref{eq:Q}) for a simulation in the strong vortex regime with $Re_\Gamma=1000$, $I_P=0.05$. Regions in blue indicate regions of strongly positive $Q$. The wire is shown as a solid black object.}
    \label{fig:q-re1000-ip0p05-tube}
\end{figure}

Figure \ref{fig:q-re1000-ip0p05-tube} depicts a volume rendering of $Q$ for a simulation within the strong vortex regime ($Re_\Gamma=1000$, $I_P=0.05$). As discussed in Ref.~\cite{krishnamoorthy1998three} and others, the vortex ``pre-senses'' the wire's presence prior to its approach to the core. Even before the wire penetrates the core, the central area of the vortex is disrupted. As the wire gradually traverses the core, numerous secondary structures coil around the primary tube— a detailed analysis of these occurs later during vorticity visualization. By the end of the interaction, little remains of the tube the wire passed through. The BVI process has significantly diminished the vortex's strength by its conclusion.

\begin{figure}
    \centering
    \includegraphics[width=\textwidth]{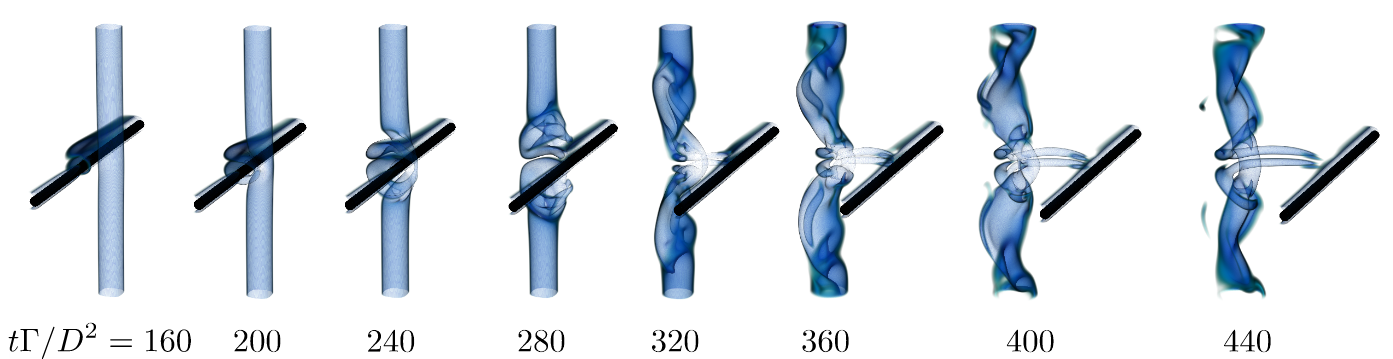}
    \caption{Volume rendering visualization of the time evolution of $Q$ (Eq.~\ref{eq:Q}) for a simulation in the weak vortex regime with $Re_\Gamma=1000$, $I_P=0.25$. Regions in blue indicate regions of strongly positive $Q$. The wire is shown as a solid black object.}
    \label{fig:q-re1000-ip0p25-tube}
\end{figure}

Figure \ref{fig:q-re1000-ip0p25-tube} depicts a similar volume rendering of $Q$ in the weak vortex regime ($Re_\Gamma=1000$, $I_P=0.25$). In this regime, the primary vortex does not ``pre-sense'' the wire, instead interacting with it solely when it is in close proximity to the vortex core. The vortex swiftly traverses the tube, causing significant deformation to the core. This deformation generates upwards and downwards-propagating waves. As time progresses, the wire completes its passage through the primary vortex tube, leaving behind a minor wake and a distorted tube. Surprisingly, despite the distortion, the ``weak'' vortex tube displays better resilience compared to the scenario involving a strong vortex. By the end of the BVI process, the waves have transversed the tube and a distorted vortex tube remains behind. Counter-intuitively this means that at $Re_\Gamma=1000$, a vortex tube endures the ``cut'' of its core more effectively when encountering a weaker cutting object. Primarily, this outcome emerges because the interaction with a slower wire is prolonged, allowing the vortex more time to engage with secondary vorticity originating from the wire. This extended interaction disrupts the primary vortex to a greater extent than the short interaction of the weak vortex regime.

\begin{figure}
    \centering
    \includegraphics[width=0.8\textwidth]{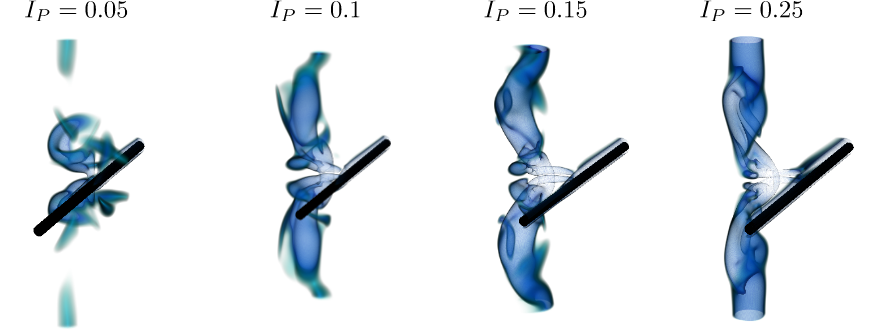}
    \caption{Volume rendering visualization of $Q$ (Eq.~\ref{eq:Q}) after the wire has passed through the cylinder axis for $Re_\Gamma=1000$, and $I_P=0.05$, $0.1$, $0.15$ and $0.25$ (left to right). Regions in blue indicate regions of strongly positive $Q$. The wire is shown as a solid black object. }
    \label{fig:q-re1000-laststage}
\end{figure}

Figure \ref{fig:q-re1000-laststage} highlights this result by showing $Q$ visualizations for the four simulated cases at various $I_P$ values after the wire's passage. In the case of the smallest $I_P$, the vortex experiences complete disruption. However, for $I_P=0.15$ and $0.25$, the vortex undergoes deformation, exhibiting propagating waves along its axis, yet retains some coherence. The $I_P=0.1$ scenario lies between these extremes, displaying weakening in certain regions and deformation in others. Notably, the extent and nature of disruption in the primary core serve as markers distinguishing between weak and strong vortex regimes. The behavior observed in Figure \ref{fig:q-re1000-laststage} suggests a smooth transition between these regimes taking place around $I_P\approx0.1$. This was observed in experiments, and a value of $I_P\approx 0.08$ was given for the transition between regimes in Ref.~\cite{krishnamoorthy1998three}. 

\subsection{The strong vortex regime ($I_P=0.05)$}
\label{sec:strongv}

As the weak and strong vortex regimes show very different dynamics, they will be discussed separately, taking the lowest and highest values of $I_P$ as paradigmatic for each regime. This subsection discusses the strong vortex regime, with $I_P=0.05$. 

\begin{figure}
    \centering
    \includegraphics[width=\textwidth]{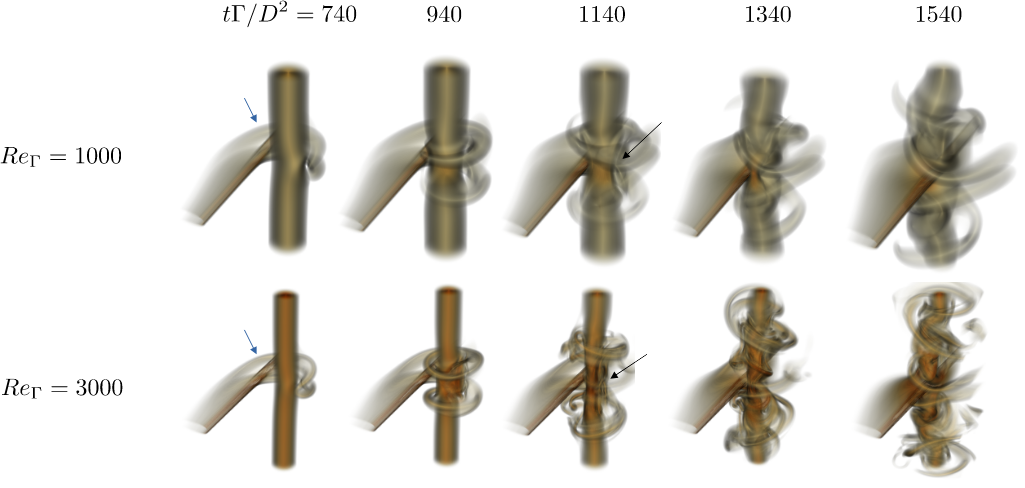}
    \caption{Volume rendering visualization of the time evolution of $|\omega|$ during the early stages for a simulation in the strong vortex regime $I_P=0.05$ with $Re_\Gamma=1000$ (top row) and $Re_\Gamma=3000$ (bottom row). }
    \label{fig:ww-ip0p05-early}
\end{figure}

To investigate the role of secondary vorticity in the vortex-wire interaction, Figure \ref{fig:ww-ip0p05-early} presents volumetric representations of the vorticity magnitude $|\omega|$ at the early stages of interaction for both $Re_\Gamma=1000$ and $Re_\Gamma=3000$. This stage of the interaction is rich in details, and the visualizations align with the experimental images and descriptions in Ref.~\cite{krishnamoorthy1998three}: as the vortex approaches the tube, vorticity from its boundary layer detaches, and begins to curl around the primary tube. This is marked with a blue arrow at $t\Gamma/D^2=740$. Even before the wire approaches the vortex core, this detached vorticity loops once around the main vortex,  distorting the core, as seen at $t\Gamma/D^2=940$. This is because the wrapped vorticity induces velocities which commence a compression or ``necking'' of the core. This phenomena is accentuated as the cylinder approaches the primary vortex. In the impact zone, the vortex and cylinder have oppositely signed vorticity, which, when pressed together further thins the neck region. The primary vorticity increases to conserve circulation, and a region with increased vorticity magnitude becomes apparent at $t\Gamma/D^2=1140$ (marked with a black arrow) and $t\Gamma/D^2=1340$. At the same time, the detached vorticity propagates further along the vortex axis, nearly reaching the computational domain's periodic boundaries as seen in the last snapshot at $t\Gamma/D^2=1540$. 

The observations detailed here mirror those documented experimentally in Ref.~\cite{krishnamoorthy1998three}, while expanding upon the experimental groundwork in several crucial ways. Firstly, Figure \ref{fig:ww-ip0p05-early} portrays vorticity three-dimensionally rather than photographs of dye visualization. This representation enables us to illustrate the propagation of secondary vorticity through the axis, a phenomenon which is not clear in the experiments, as well as showing the new vortex filaments produced. Secondly, we demonstrate that secondary vorticity arises without the need to introduce a source of turbulence, either through background flow or vibrations of the structure. Even starting from simple initial conditions, such as a Lamb-Oseen vortex, the secondary vorticity originating solely from the wire's boundary layer is enough to faithfully reproduce the significant disruptions in the primary vortex seen experimentally. Lastly, we note that our results are roughly symmetrical around the cutting plane in our results due to the absence of axial flow. This shows that the presence of a weak axial flow does not change the phenomenology of strong vortex BVI very much. We also note that for both Reynolds numbers, the flow phenomenology at the early stages is very similar, including the extent of wrapping and speed of propagation of secondary vorticity around the primary vortex, even if the vorticity itself appears much sharper due to the reduced viscosity. This facilitates the comparison to the experiment.

\begin{figure}
    \centering
    \includegraphics[width=0.32\textwidth]{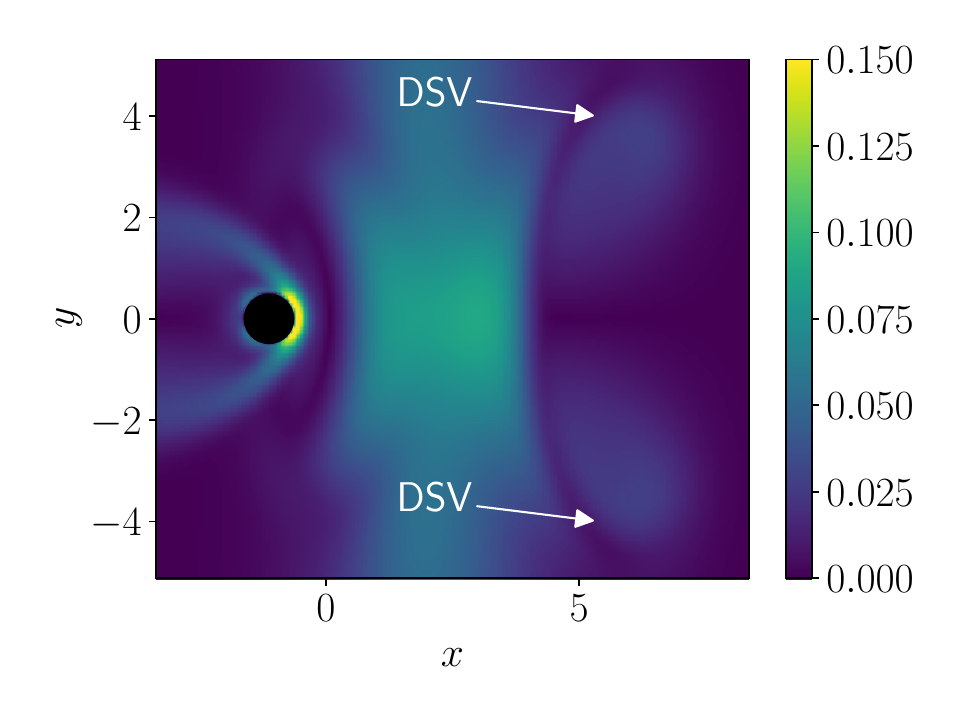}
    \includegraphics[width=0.32\textwidth]{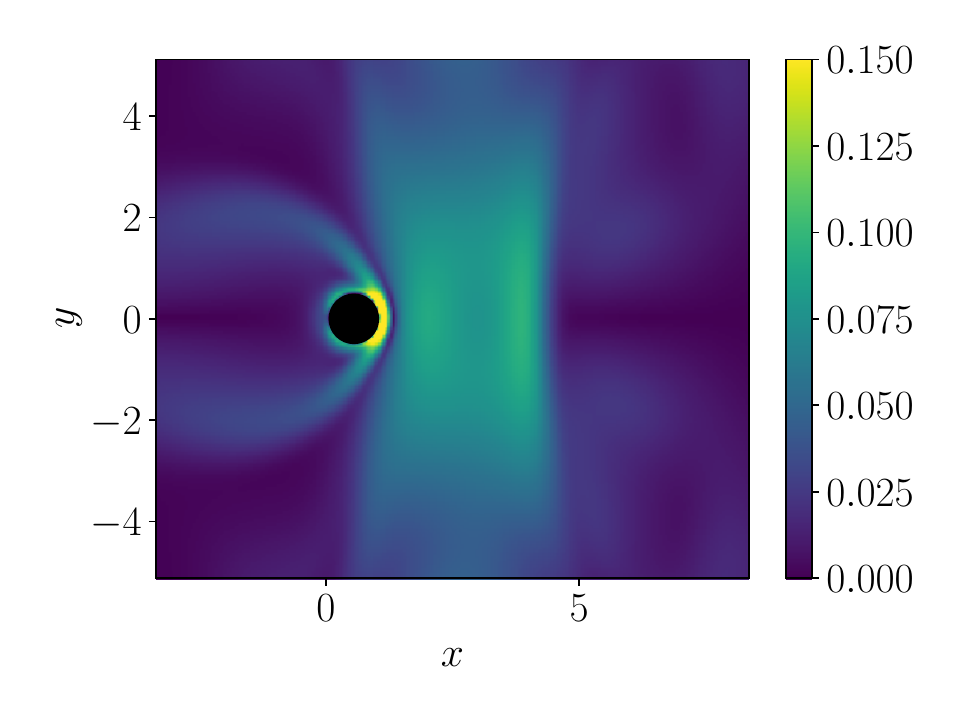}
    \includegraphics[width=0.32\textwidth]{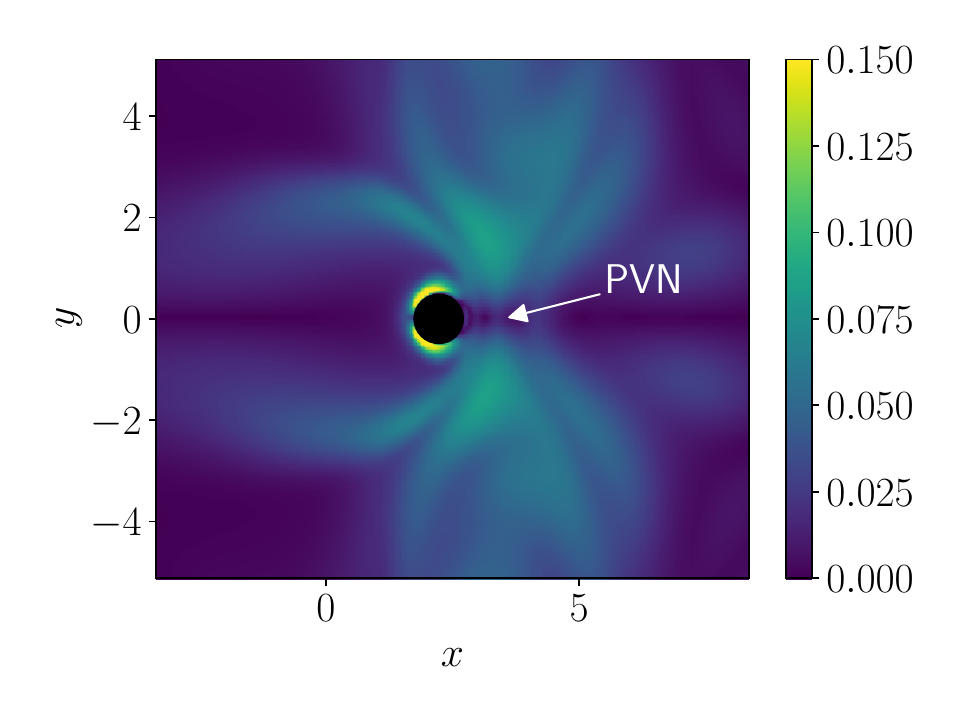}\\
    \includegraphics[width=0.32\textwidth]{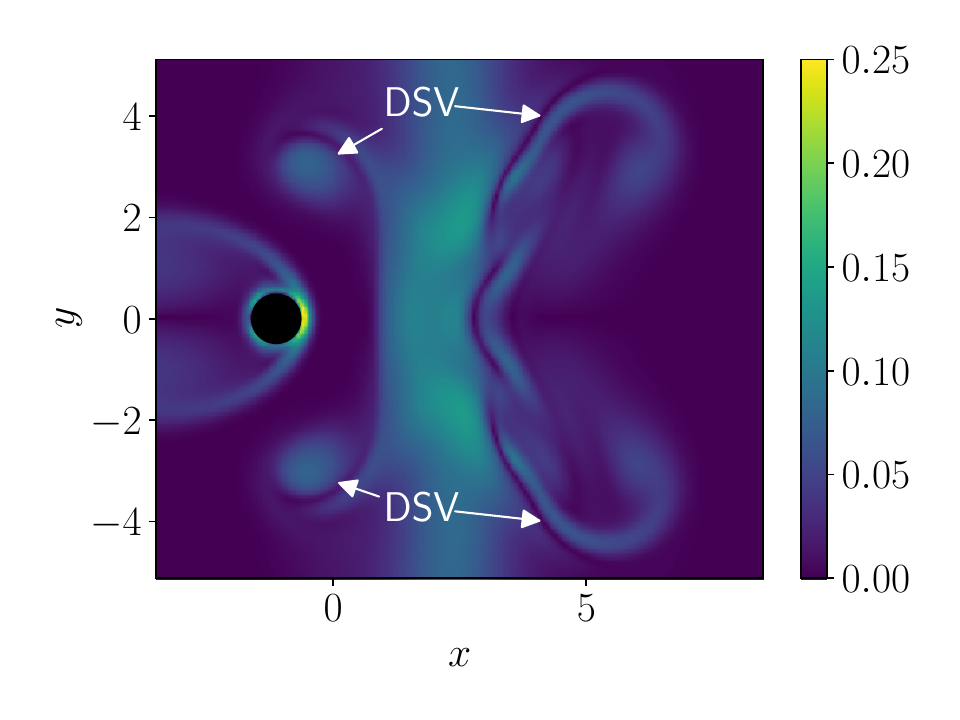}
    \includegraphics[width=0.32\textwidth]{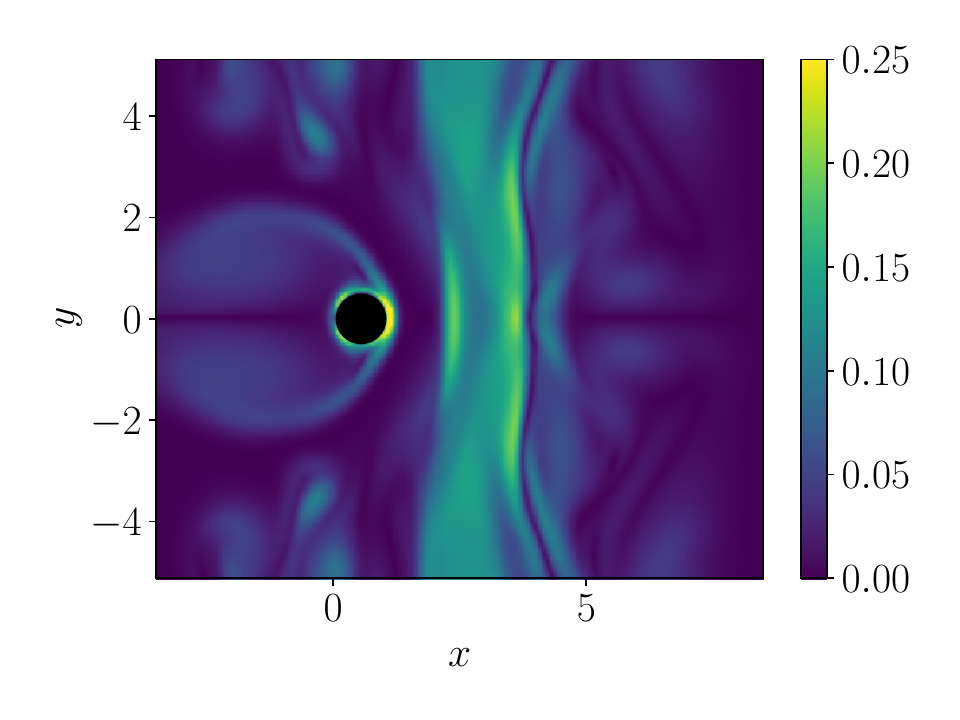}
    \includegraphics[width=0.32\textwidth]{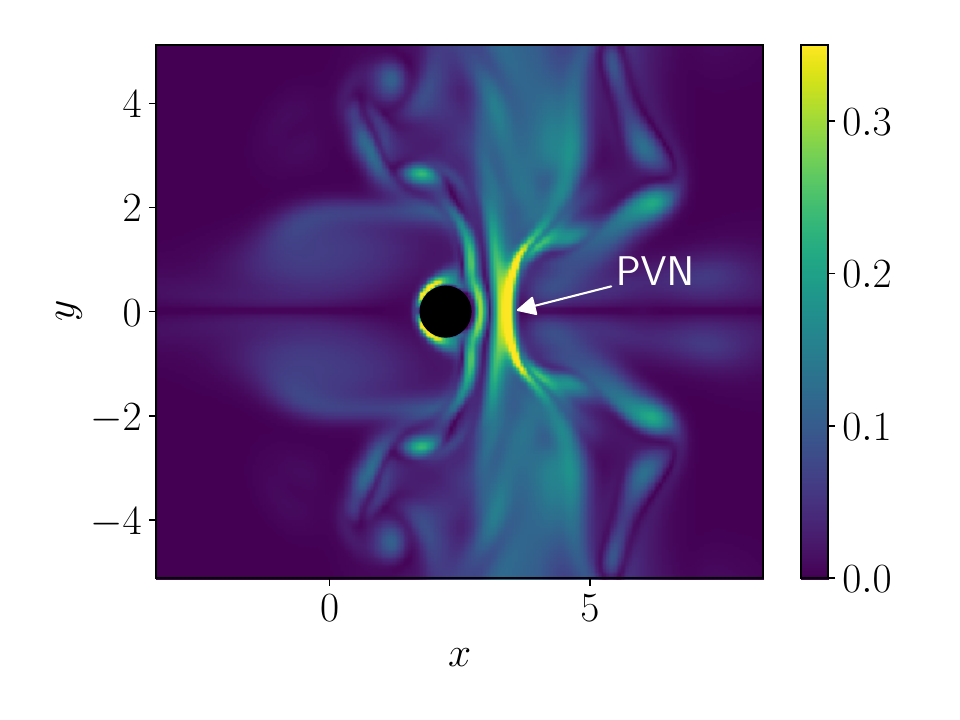}
    \caption{Vorticity magnitude $|\omega|$ at the $z=0$ plane evolution just before impact in the strong vortex regime with $I_P=0.05$ at $Re_\Gamma=1000$ (top row) and $Re_\Gamma=3000$ (bottom row). Snapshots are at $t\Gamma/D^2=940$ (left), $1140$ (middle) and $1340$ (right). Abbreviations: ``DSV'' detached secondary vorticity, ``PVN'' primary vortex necking.}
    \label{fig:ww-zcut-ip0p05}
\end{figure}

To focus on the early stages of the collision, Figure \ref{fig:ww-zcut-ip0p05} provides more qualitative data by representing the vorticity magnitude plotted at the central $x$-$y$ plane at three time instants as the wire is approaching the tube. The early detachment of secondary vorticity is seen in the left panels of Figure \ref{fig:ww-zcut-ip0p05}. Significant detachment and wrapping are present even if the wire is several diameters away from the core. The detached vorticity appears much sharper at $Re_\Gamma=3000$, but the degree of wrapping is similar, as will be confirmed below. The middle panels show how the primary vortex has already been distorted before the wire begins to penetrate into the vortex. The wrapped vorticity induces velocities which begin to thin out the center of the vortex. This process is accentuated as the physical interaction begins- the core is further compressed producing a ``necking'' seen in the right panels. We note that at $Re_\Gamma=1000$ even if the core is compressed, the vorticity intensification is smaller (or non-present). The reasons for this will be analyzed below.

\begin{figure}
    \centering
    \includegraphics[width=0.32\textwidth]{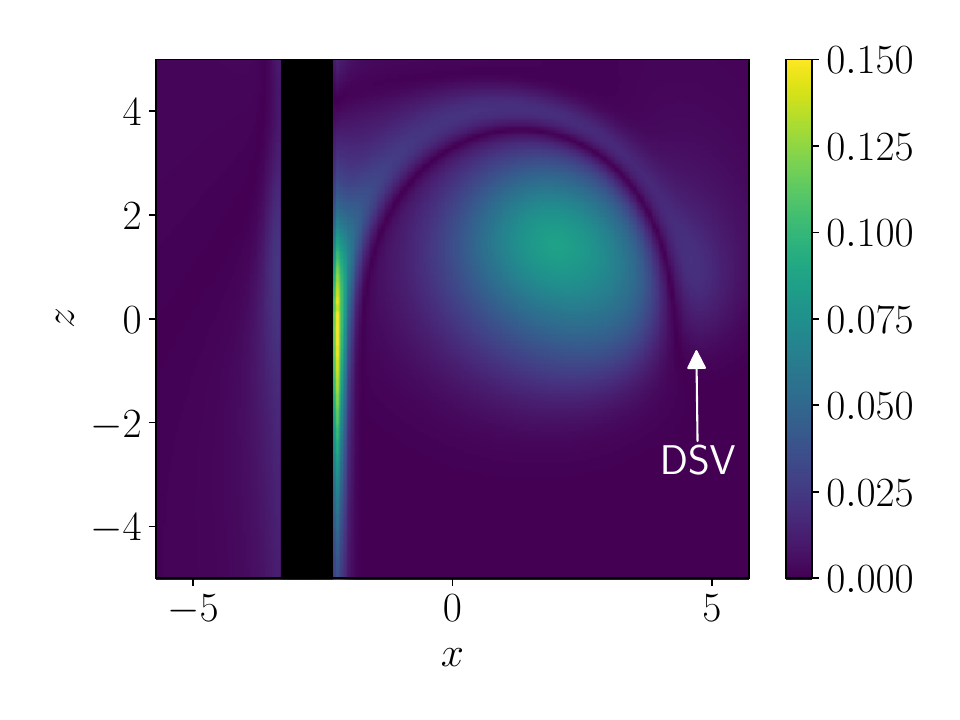}
    \includegraphics[width=0.32\textwidth]{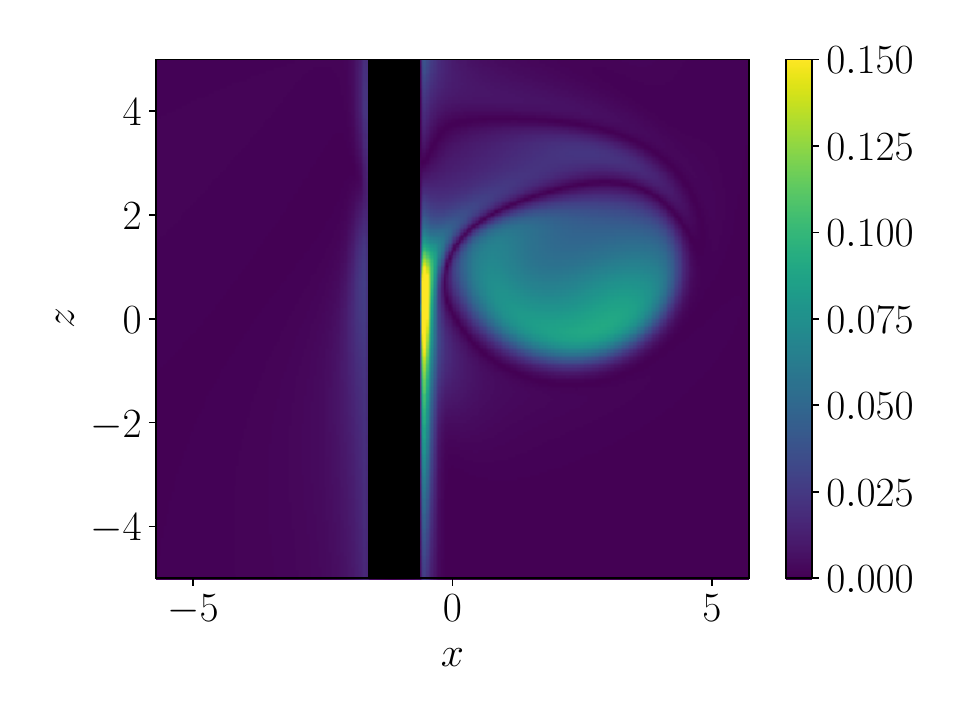}
    \includegraphics[width=0.32\textwidth]{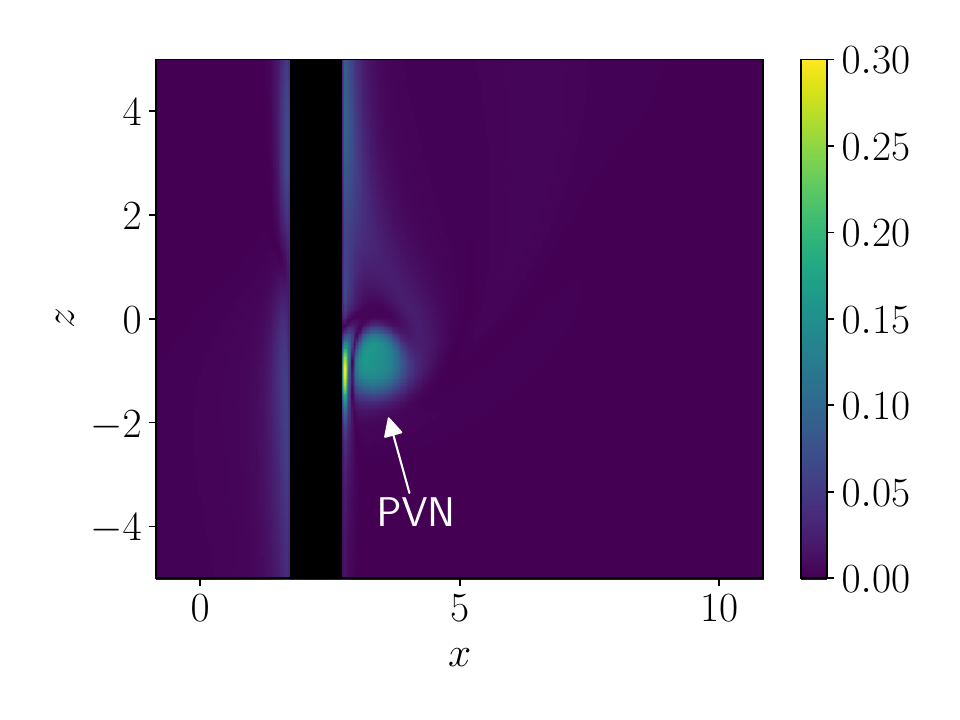}\\
    \includegraphics[width=0.32\textwidth]{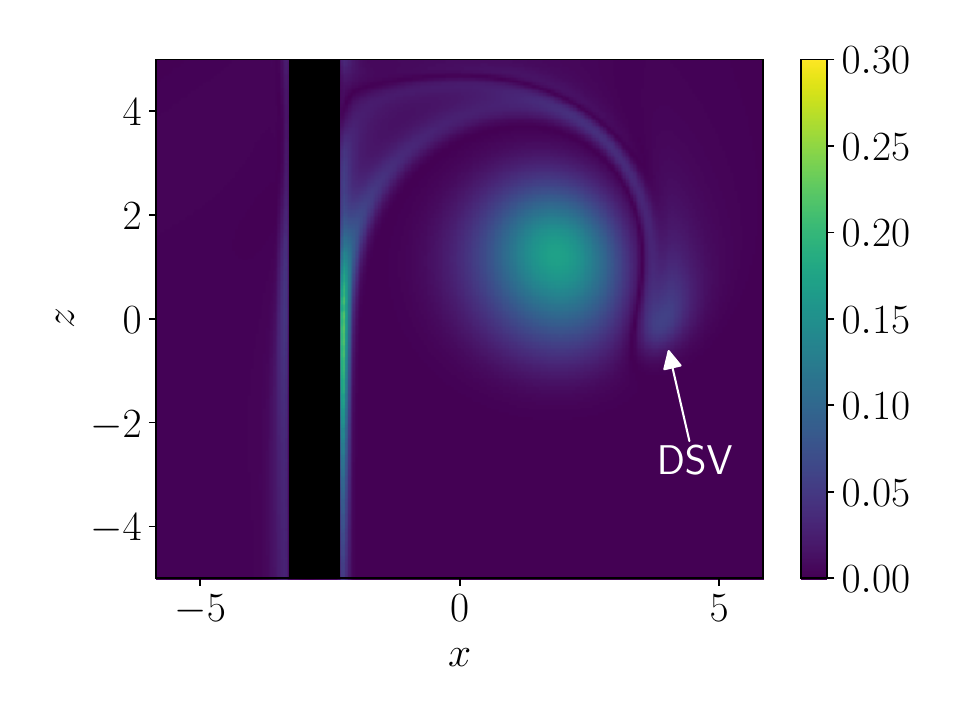}
    \includegraphics[width=0.32\textwidth]{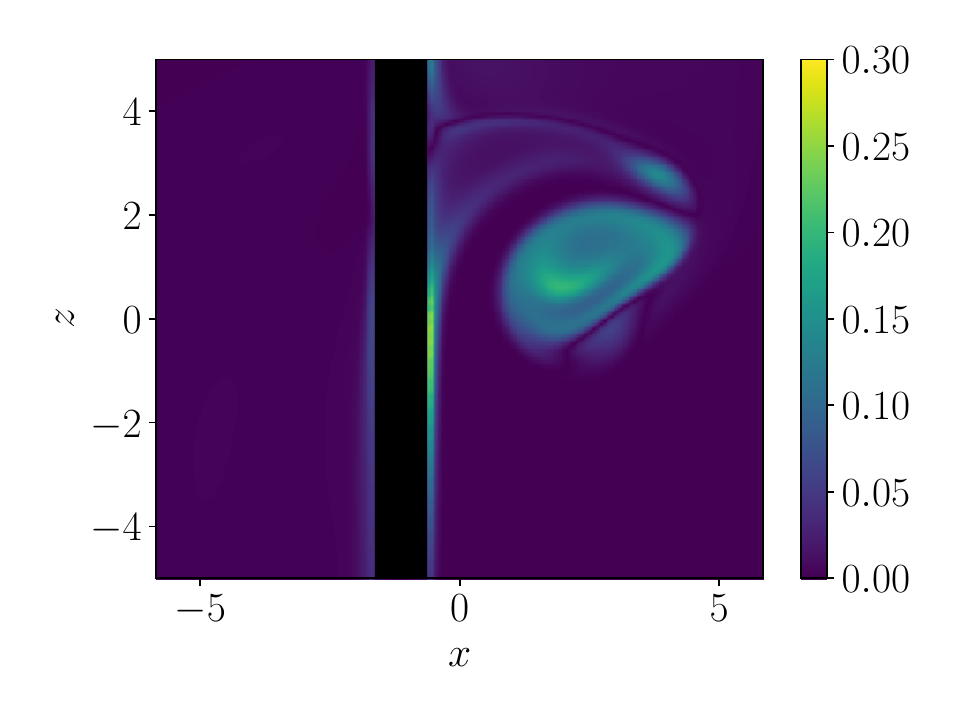}
    \includegraphics[width=0.32\textwidth]{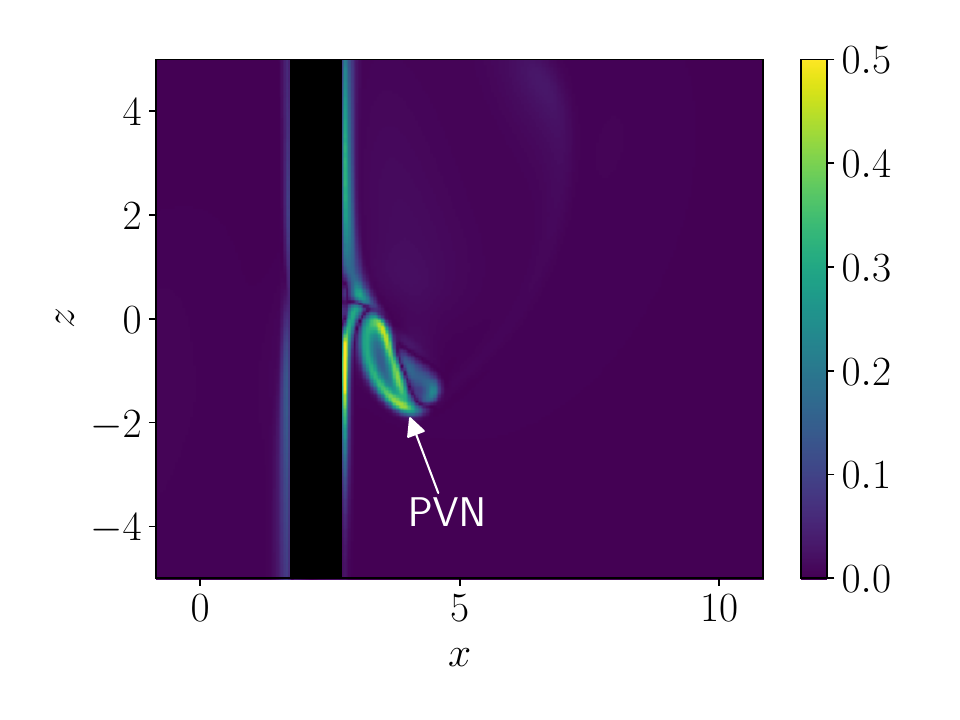}
    \caption{Vorticity magnitude $|\omega|$ at the $y=0$ mid-plane evolution for a simulation in the strong vortex regime at $I_P=0.05$ with $Re_\Gamma=1000$ (top row) and with $Re_\Gamma=3000$ (bottom row). Snapshots are at $t\Gamma/D^2=740$ (left), $940$ (middle), $1340$ (right). Abbreviations: ``DSV'' detached secondary vorticity, ``PVN'' primary vortex necking. }
    \label{fig:ww-ycut-ip0p05-early}
\end{figure}

Another view of the process is shown in Figure \ref{fig:ww-ycut-ip0p05-early} which presents the vorticity magnitude plotted at the central $x$-$z$ plane. The snapshots in the left column confirm that the dynamics of the secondary vorticity detachment process are approximately Reynolds independent: the degree of wrapping is the same for both Reynolds number. The middle panel also shows how the vortex is distorted by the wrapped vorticity before the close interaction begins. Finally, the right panel shows the degree of ``necking'' present in the primary vortex as the body comes close to the vortex. The middle and right panels also reveal the main effect of $Re_\Gamma$: while at $Re_\Gamma=1000$ the primary vortex remains more-or-less smooth, at $Re_\Gamma=3000$ the vorticity profile is highly deformed as vorticity concentrates in thin sheets outside the center. We also note that the core has been substantially displaced from its original position.

\begin{figure}
    \centering
    \includegraphics[width=0.8\textwidth]{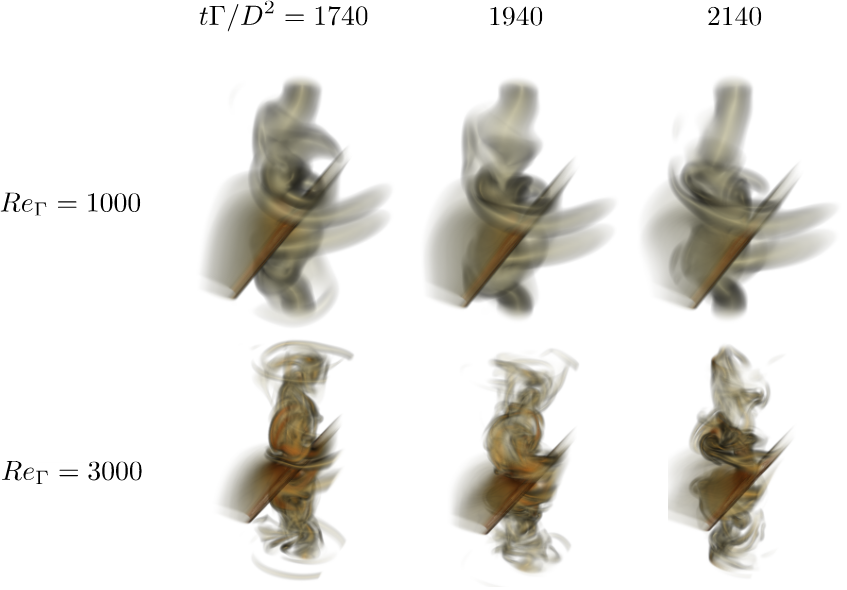}
    \caption{Volume visualization of the time evolution of $|\omega|$ during the late stages for a simulation in the strong vortex regime $I_P=0.05$ with $Re_\Gamma=1000$ (top row) and $Re_\Gamma=3000$ (bottom row). }
    \label{fig:ww-ip0p05-late}
\end{figure}

However, these differences do not result in distinct outcomes of the BVI. Figure \ref{fig:ww-ip0p05-late} shows the vorticity modulus for the late stages of the interaction, once the wire has cut the vortex core. As discussed in Section \ref{sec:transition}, in the strong vortex case the vortex remnants are much weaker than the original primary vortex. Instead, the flow appears to consist of a turbulent cloud with vortices oriented in many directions. 

\begin{figure}
    \centering
    \includegraphics[width=0.40\textwidth]{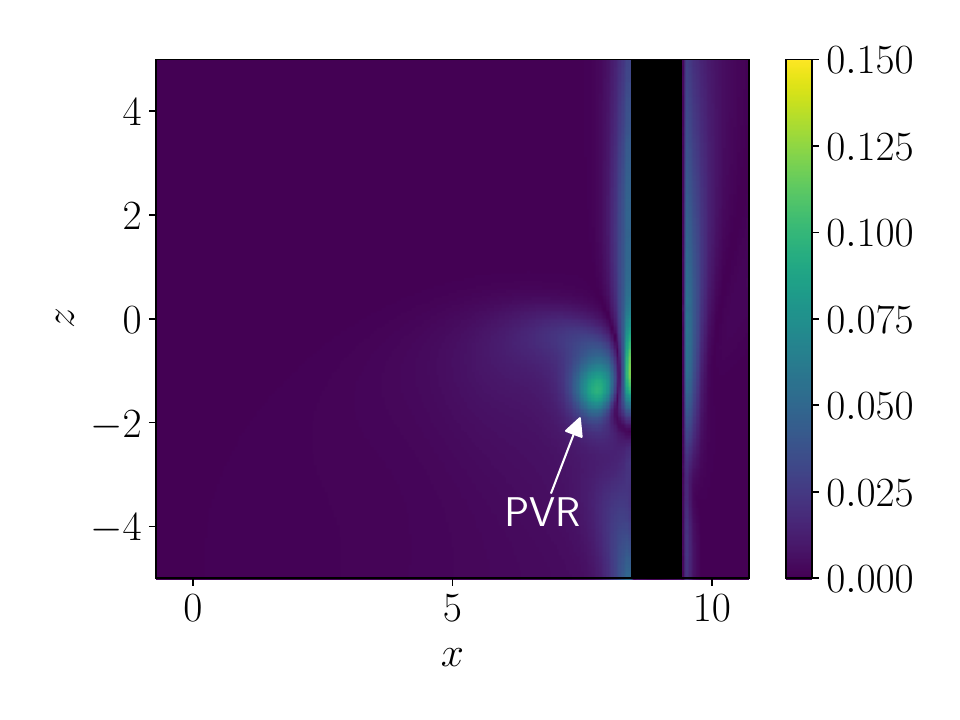}
    \includegraphics[width=0.40\textwidth]{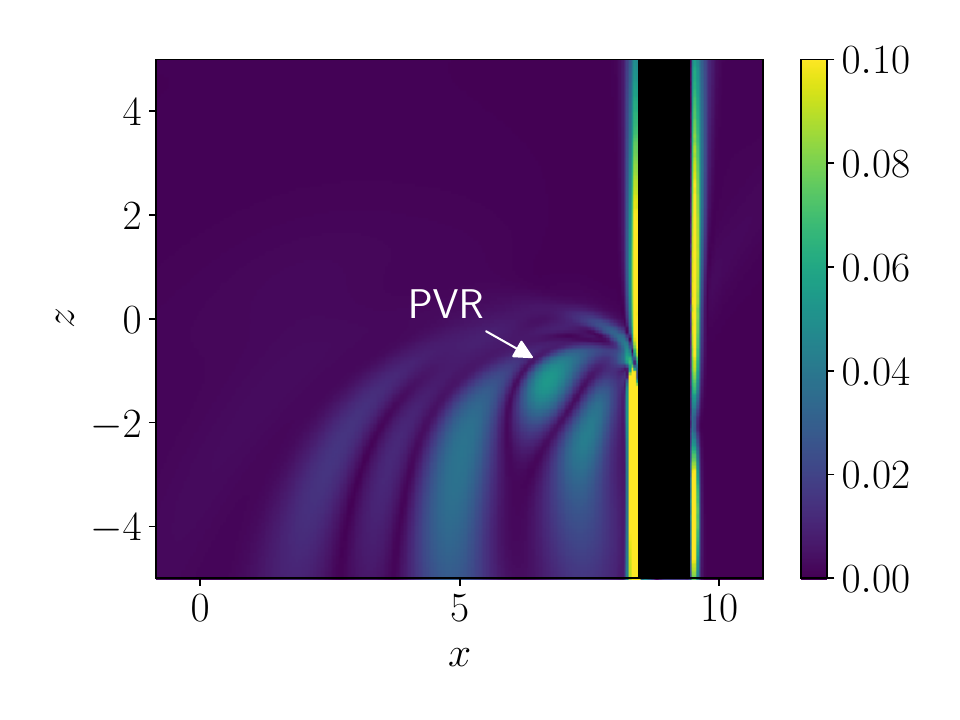}
    \caption{Vorticity magnitude $|\omega|$ at the $y=0$ mid-plane evolution and $t\Gamma/D^2=2140$ for a simulation in the strong vortex regime at $I_P=0.05$ with $Re_\Gamma=1000$ (left) and with $Re_\Gamma=3000$ (right). Abbreviations: ``PVR'' primary vortex remnants. }
    \label{fig:ww-ycut-ip0p05-late}
\end{figure}

Another view of the remnants is shown in Figure \ref{fig:ww-ycut-ip0p05-late}, which presents the vorticity at the $y=0$ plane at a late stage. For $Re_\Gamma=1000$, the remnants of the primary vortex can still be distinguished. It has been displaced, six to seven vortex radii from its initial position, and has a circulation which has been reduced to a quarter of its original strength. For $Re_\Gamma=3000$, the remnants are extremely weak, and the circulation cannot be adequately measured as a single remnant cannot be identified. In both cases, the remnants will be dissipated by viscosity.

\subsection{The weak vortex regime ($I_P=0.25$)}
\label{sec:weakv}

\begin{figure}
    \centering
    \includegraphics[width=\textwidth]{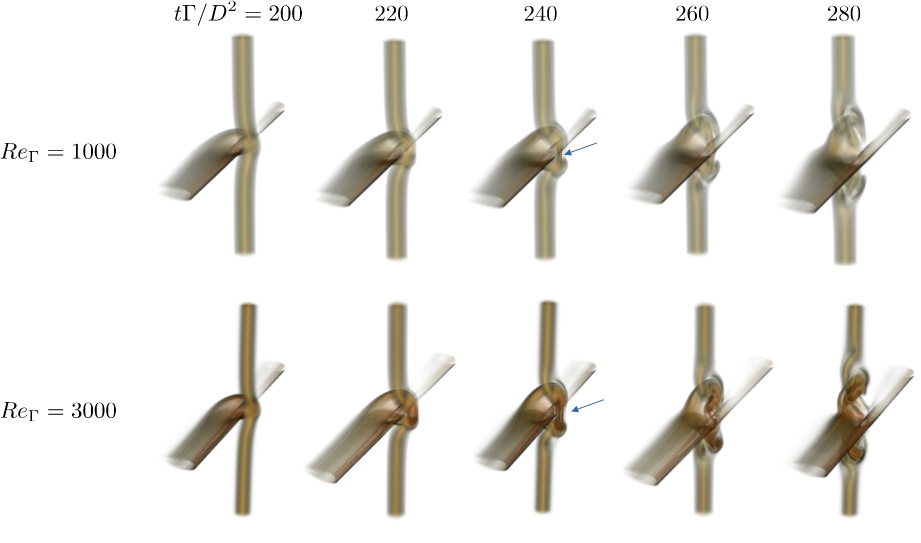}
    \caption{Volume rendering visualization of the time evolution of $|\omega|$ during the early stages for a simulation in the weak vortex regime $I_P=0.25$ with $Re_\Gamma=1000$ (top row) and $Re_\Gamma=3000$ (bottom row). }
    \label{fig:ww-ip0p25-early}
\end{figure}

Turning to the weak vortex regime, Figure \ref{fig:ww-ip0p25-early} shows volume plots of the vorticity magnitude $|\omega|$ for $I_P=0.25$ and the two Reynolds numbers simulated at the early stages of the interaction. As in the strong vortex regime, two sources of vorticity are evident: the primary vortex and the wire's boundary layer. However, the overall flow evolution looks very different when compared to Figure \ref{fig:ww-ip0p05-early}. The presence of a weaker vortex results in lower induced velocities on the wire. The wire's boundary layer is only beginning to lift off in the left-most panels at $t\Gamma/D^2=200$ and $220$, and has no time to curl around and deform the primary vortex before impact. This can be appreciated by contrasting with the visualization at $t\Gamma/D^2=740$ in Figure \ref{fig:ww-ip0p05-early} where the primary core is seen to have deformed on the right side. 

Consequently, the wire's boundary layer does not play a significant role in this scenario. This lack of interaction before the wire approaches the core serves as a key characteristic of the weak vortex regime, and has historically facilitated its modeling using inviscid methods. Instead, the physics in this regime are dominated by the displacement and deformation of the core, which is propagated through axial waves, as illustrated in the visualizations at $t\Gamma/D^2=240$ and onwards. The core is deformed due to the physical presence of the wire, which forces oppositely signed vorticity to approach the core in a process reminiscent of vortical reconnection \cite{saunders2015vorticity} (marked by a blue arrow in Figure \ref{fig:ww-ip0p25-early}). This displaces the vortex tube, and stretches it into a thin ``neck'' until the core is eventually ``cut''. This process is more marked in the $Re_\Gamma=3000$ case, where the vorticity is intensified substantially. 

Simultaneously, the core deformations close to the wire propagate through waves along the primary tube's axis, as was previously observed in Figure \ref{fig:q-re1000-ip0p25-tube}. These waves do not substantially modify the radius of the primary vortex, a phenomena observed in the simulations and experiments of weak-vortex BVI with axial flow in Ref.~\cite{marshall1997instantaneous}. Here, there is a crucial difference from the experiments with axial flow which introduces a notable asymmetry in wave propagation. Our simulations instead show a symmetrical propagation originating from the cylinder impact zone.

\begin{figure}
    \centering
    \includegraphics[width=0.32\textwidth]{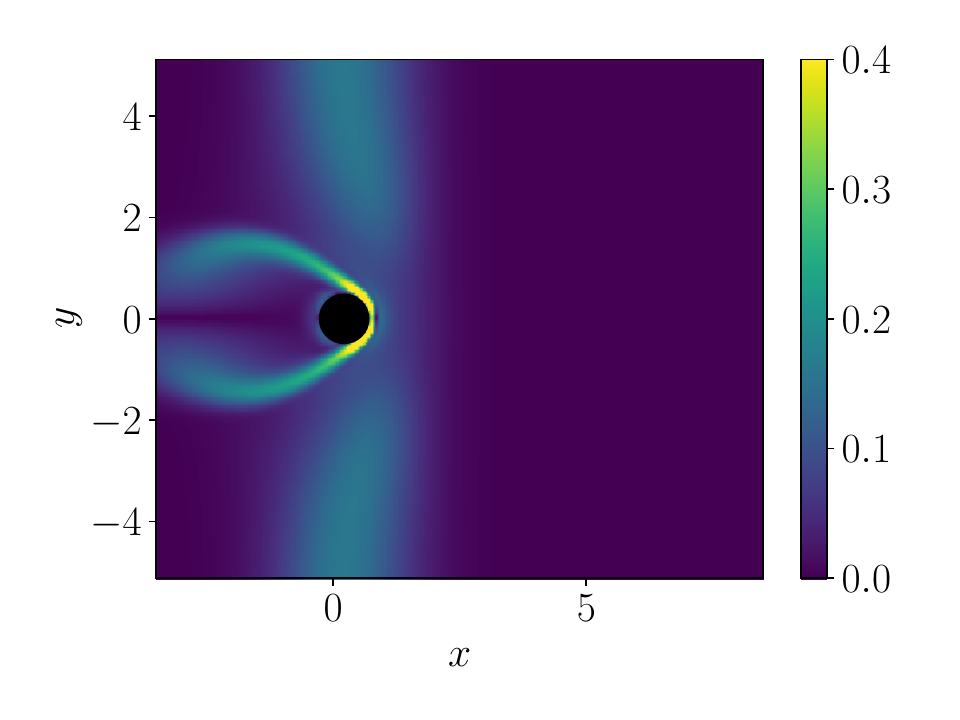}
    \includegraphics[width=0.32\textwidth]{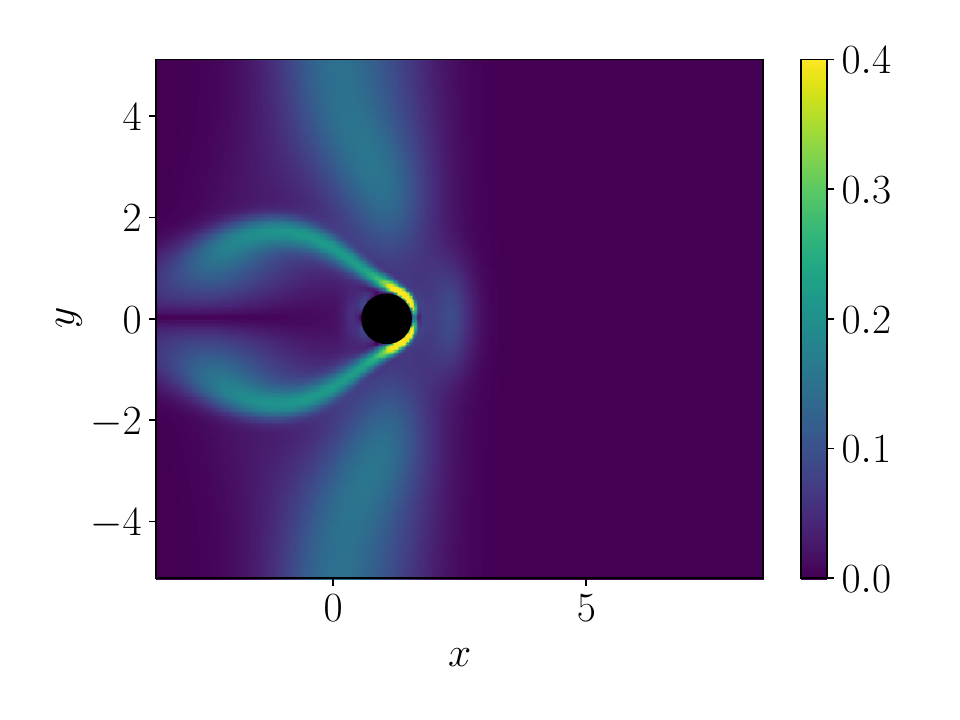}
    \includegraphics[width=0.32\textwidth]{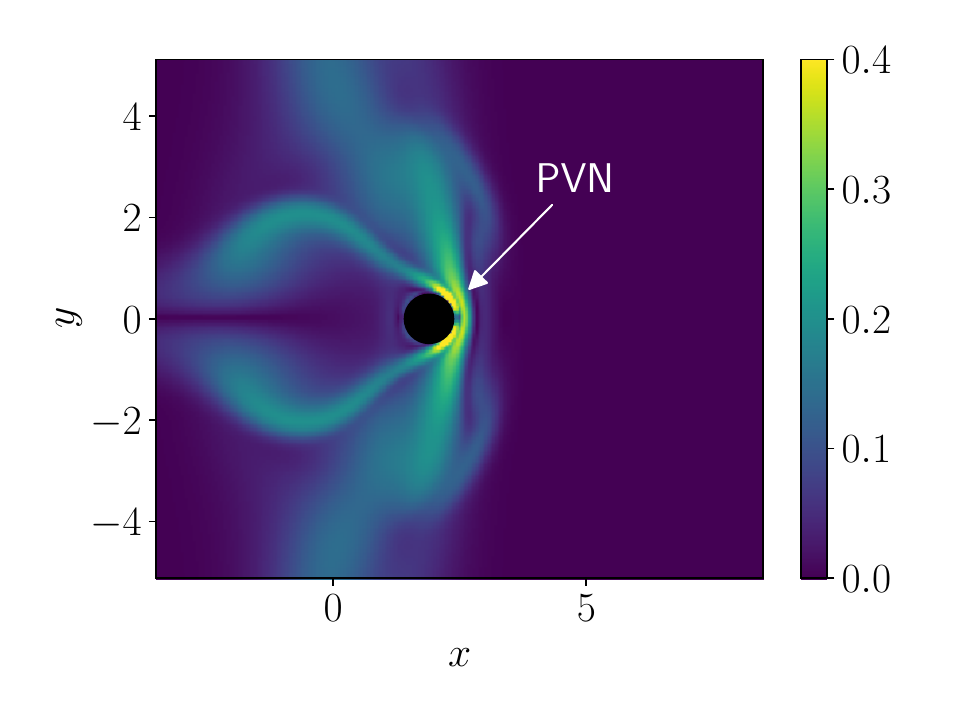}\\
    \includegraphics[width=0.32\textwidth]{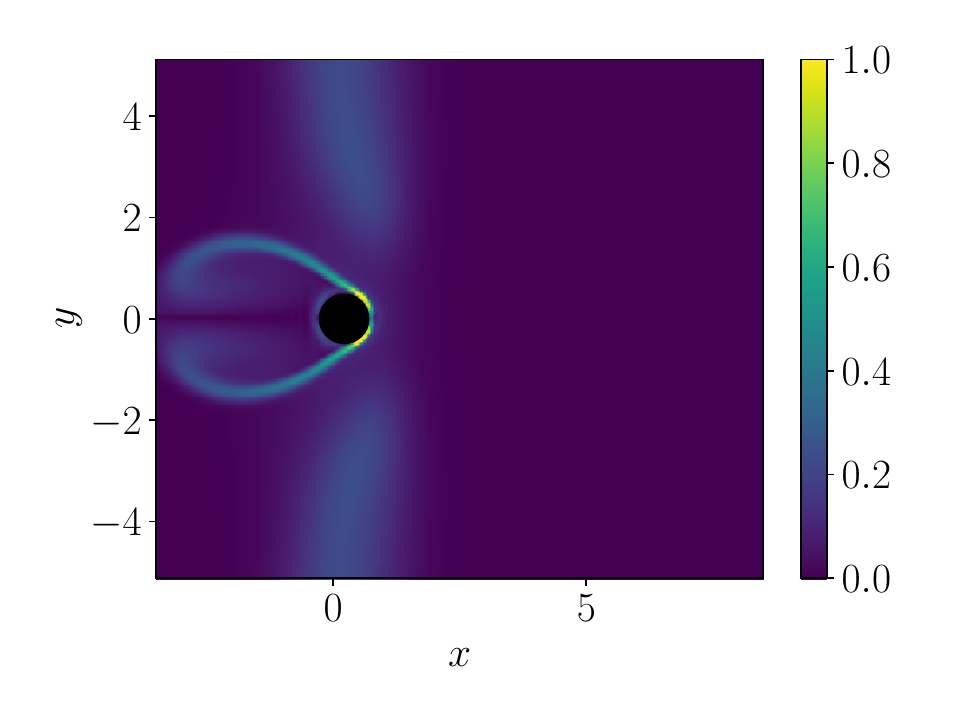}
    \includegraphics[width=0.32\textwidth]{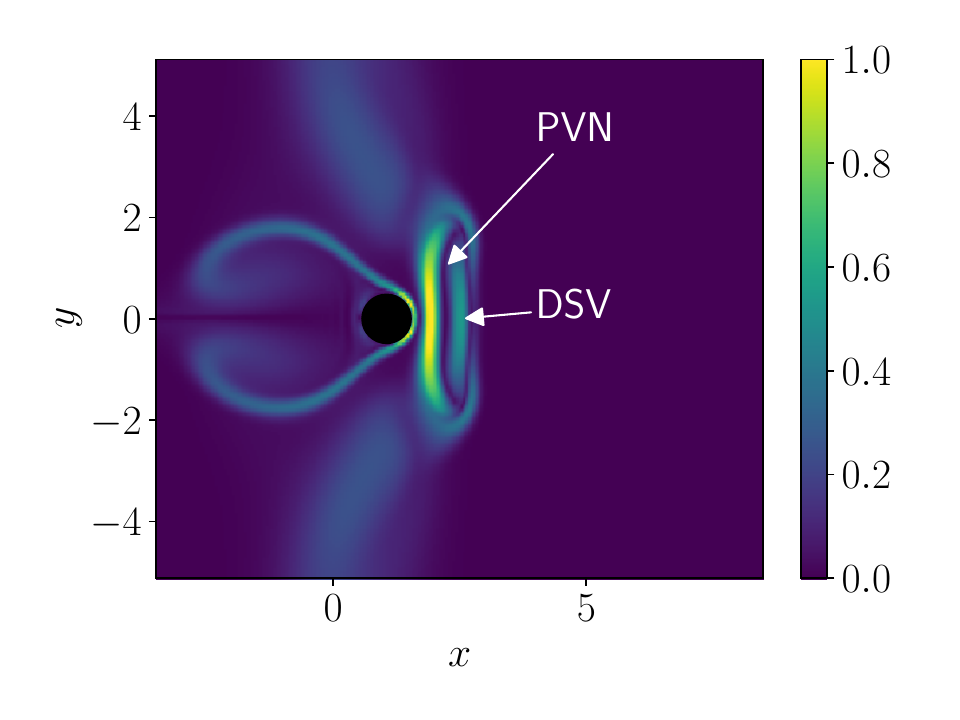}
    \includegraphics[width=0.32\textwidth]{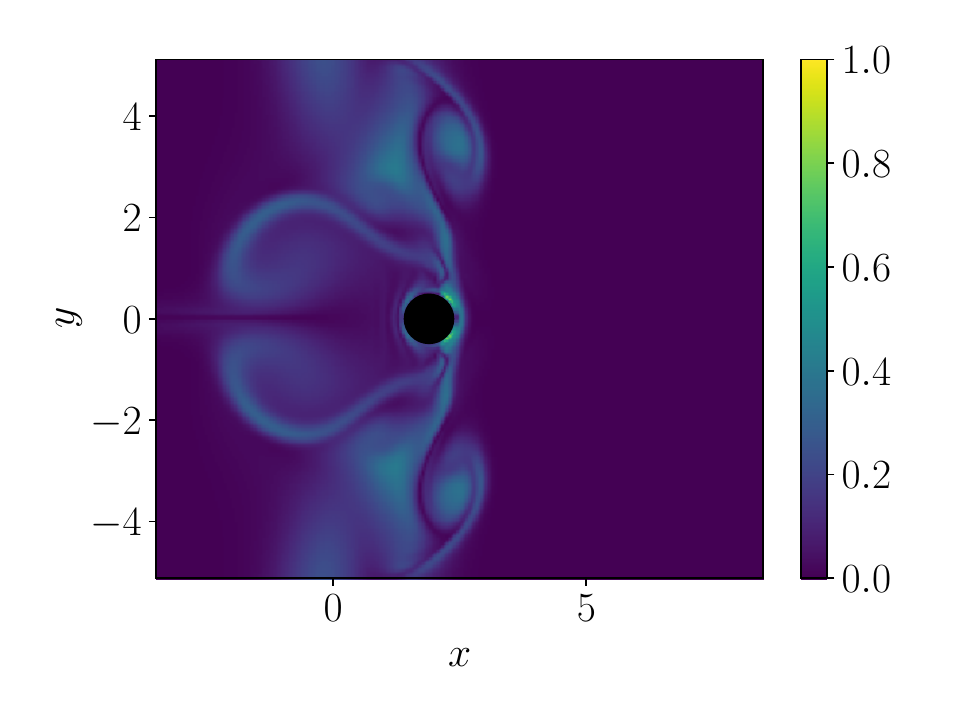}
    \caption{Vorticity magnitude $|\omega|$ at the $z=0$ plane evolution just before impact in the weak vortex regime with $I_P=0.25$ at $Re_\Gamma=1000$ (top row) and $Re_\Gamma=3000$ (bottom row). Snapshots are at $t\Gamma/D^2=220$ (left), $240$ (middle) and $260$ (right). Abbreviations: ``PVN'' primary vortex necking, ``DSV'' detached secondary vorticity.}
    \label{fig:ww-zcut-ip0p25}
\end{figure}

\begin{figure}
    \centering
    \includegraphics[width=0.32\textwidth]{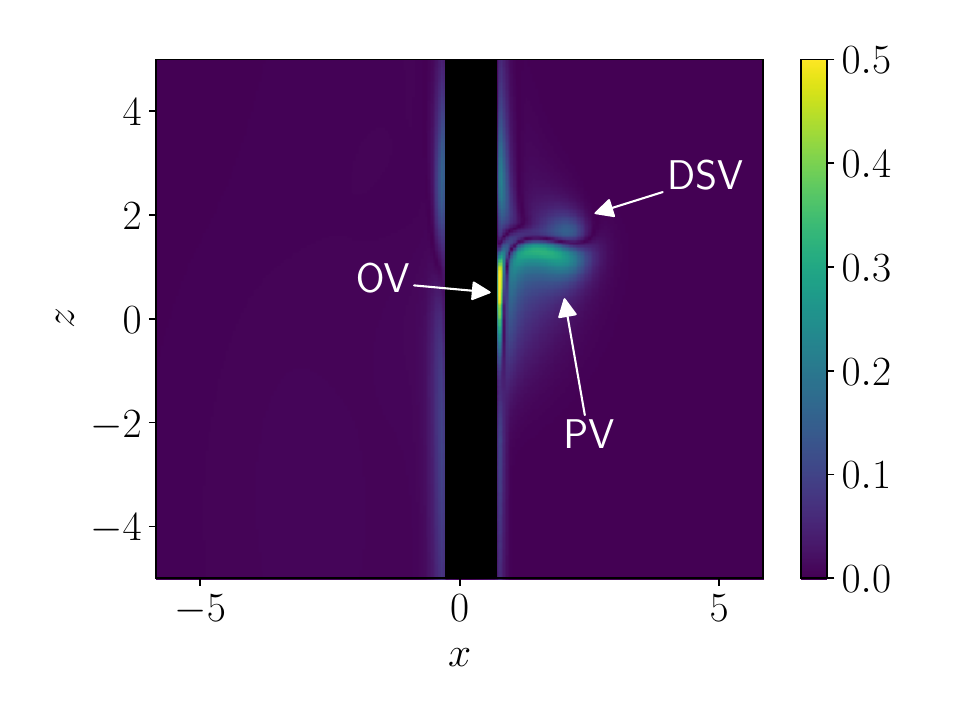}
    \includegraphics[width=0.32\textwidth]{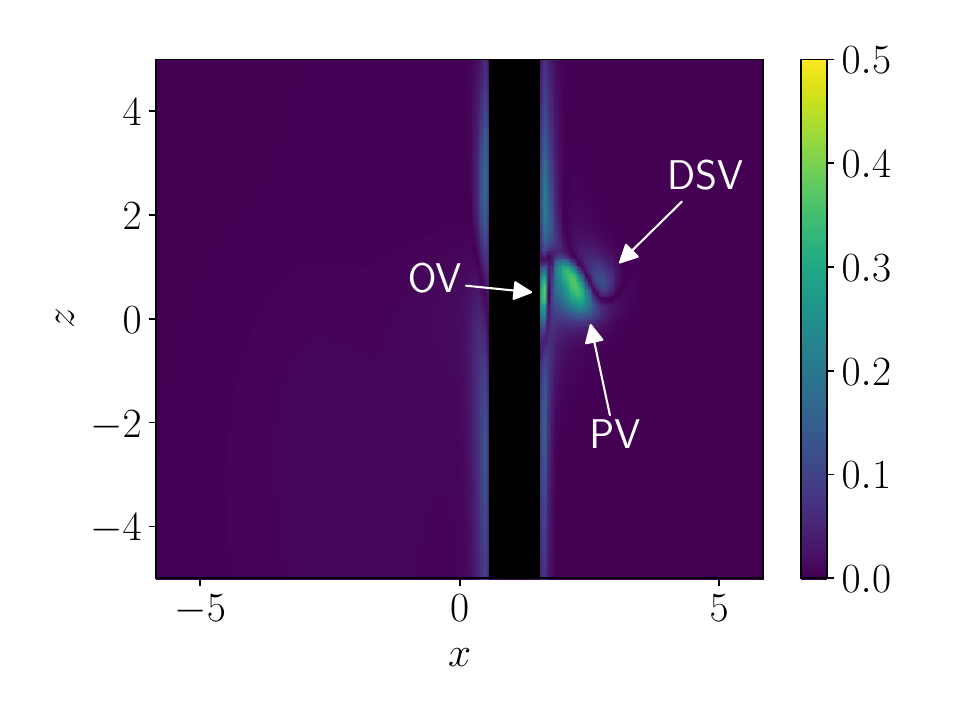}
    \includegraphics[width=0.32\textwidth]{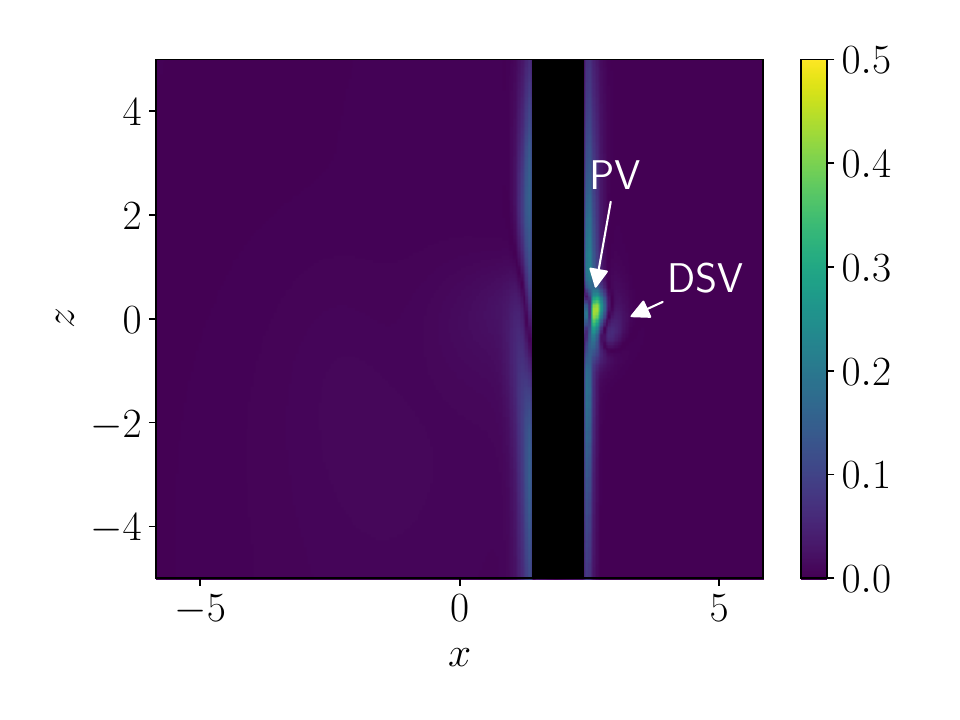}\\
    \includegraphics[width=0.32\textwidth]{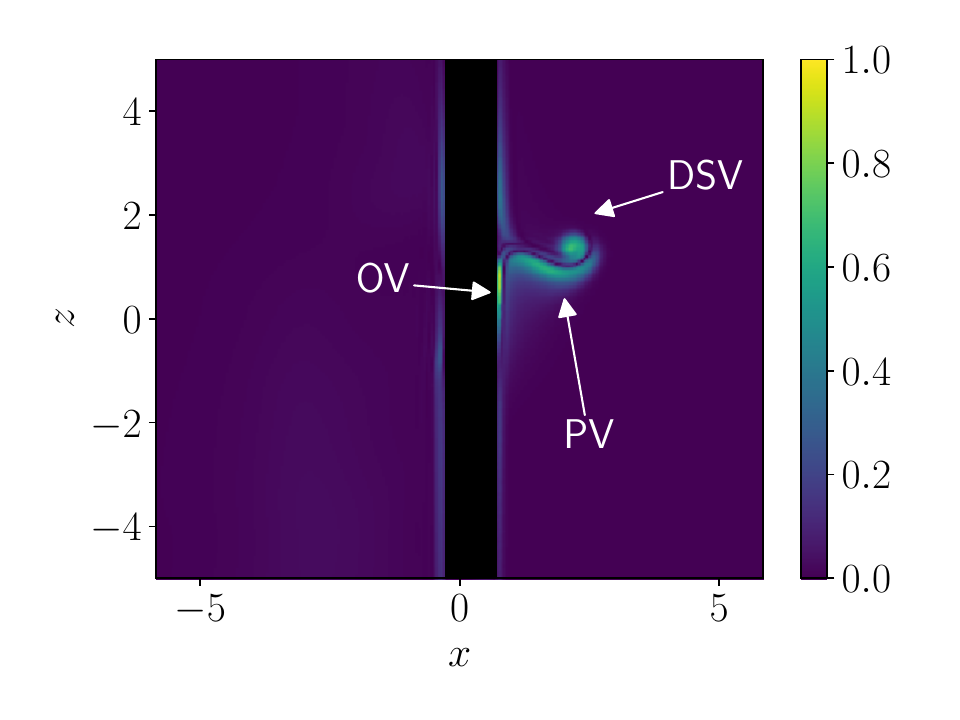}
    \includegraphics[width=0.32\textwidth]{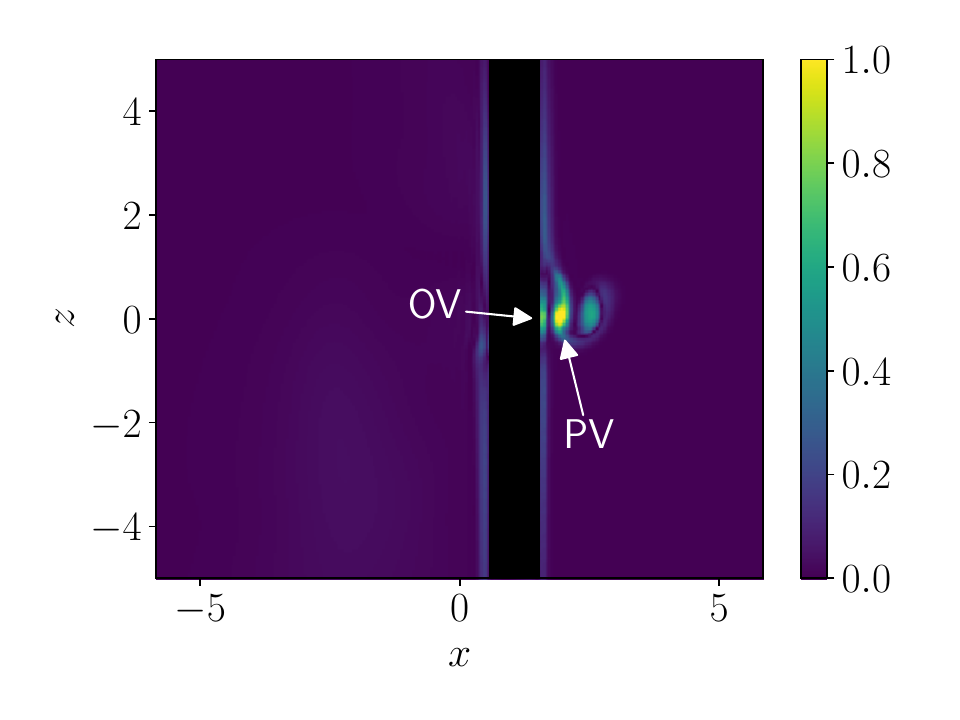}
    \includegraphics[width=0.32\textwidth]{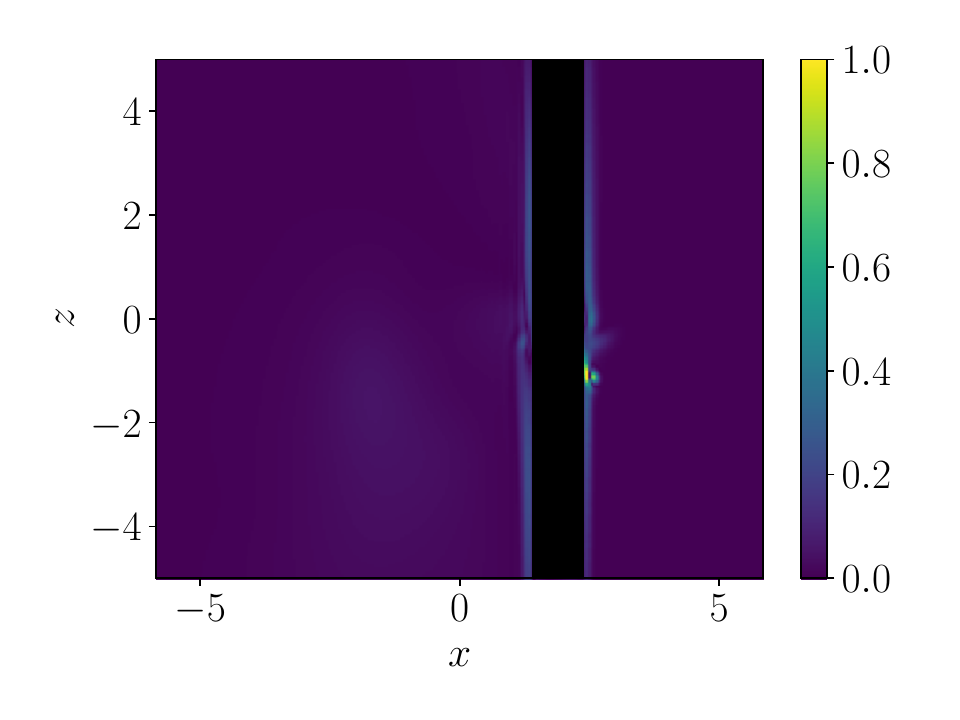}\\
    \caption{Vorticity magnitude $|\omega|$ at the $y=0$ mid-plane evolution for a simulation in the weak vortex regime at $I_P=0.25$ with $Re_\Gamma=1000$ (top row) and with $Re_\Gamma=3000$ (bottom row). Snapshots are at $t\Gamma/D^2=220$ (left), $240$ (middle), $260$ (right). Abbreviations: ``DSV'' detached secondary vorticity, ``OV'' opposite vorticity, ``PV'' primary vortex. }
    \label{fig:ww-ycut-ip0p25}
\end{figure}

As in the previous section, a closer look is provided by looking at two-dimensional cuts. Figure \ref{fig:ww-zcut-ip0p25} represents the vorticity magnitude plotted at the central $x$-$y$ plane at three time instants as the wire is approaching the tube for the two Reynolds numbers, and Figure \ref{fig:ww-ycut-ip0p25} shows the vorticity through the $x$-$z$ plane for the same time instants and the two Reynolds numbers. 

The Reynolds number dependence is first reflected as thinner and more intense boundary layers. For the first panel at $t\Gamma/D^2=220$, the secondary vorticity has not had time to completely wrap around the core, and is not visible in the $z=0$ cut. The primary vortex has been deformed by the presence of the wire which is pushing oppositely signed vorticity towards it. Notably, although the vorticity has less time to detach, it exhibits a higher value than in Figure \ref{fig:ww-ycut-ip0p05-early}. This can be attributed to the non-dimensionalization using $\Gamma/\sigma$ as a velocity scale, where higher values of $I_P$ lead to greater wire velocities, resulting in larger vorticity values for a boundary layer when normalized using $\Gamma$-$\sigma$ units. 

By $t\Gamma/D^2=240$, the secondary vorticity has had time to begin to wrap around the primary vortex. As suggested by Ref.~\cite{marshall1997instantaneous}, the primary vortex can entrain a small quantity of vorticity from the boundary layer during the interaction. However, it is almost indiscernible for the panel at $Re_\Gamma=1000$, and for $Re_\Gamma=3000$, while visible, it is eclipsed by the vorticity intensification due to the core stretching and deformation. The intensification of vorticity is particularly marked for the higher Reynolds number, a phenomena which is again reminiscent of reconnection, where increases in Reynolds number produce increasingly thinner vortex sheets of large vorticity magnitudes \cite{hussain2011mechanics}. However, once viscosity takes over, annihilating or reconnecting the primary and opposite vorticity, this does not seem to have a bearing on the rest of the flow. This can explain why flow models which neglect secondary vorticity can adequately capture the interaction within this regime: the phenomenology is primarily driven by the deformations induced by the moving object, rather than the annihilation of vorticity. 

By the latest time, $t\Gamma/D^2=260$, the cylinder is cutting the vortex, and almost no vorticity is visible in the plots close to the wire. By returning to the three-dimensional visualization of Figure \ref{fig:ww-ip0p25-early}, it is clear that the cut has taken place. In these simulations, some of the phenomena usually attributed to the bending regime can be observed, i.e.~deformations of the primary vortex, and the stretching (or ``necking'') of the core near the impact location. This hints at the fact that the transition between strong vortex and bending regimes, which is outside the scope of this paper, will also be smooth.

\begin{figure}
    \centering
    \includegraphics[width=\textwidth]{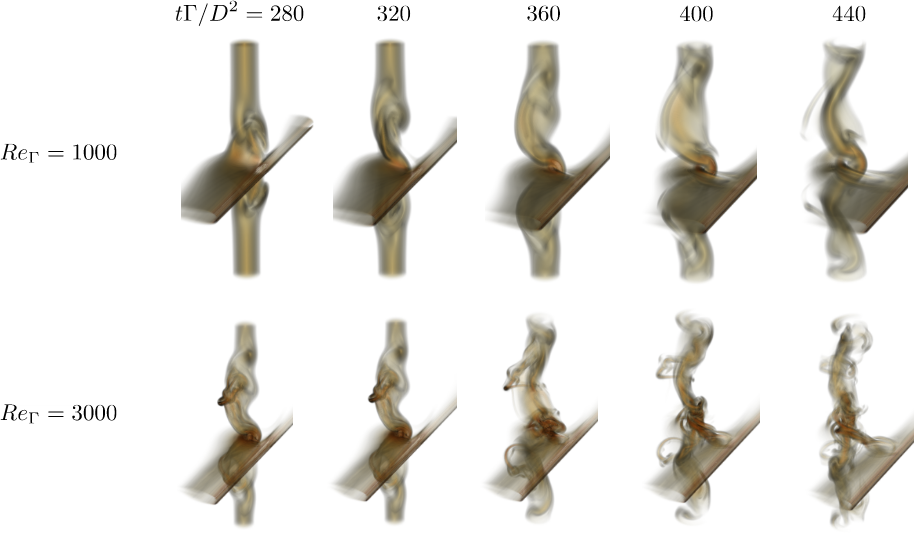}
    \caption{Volume visualization of the time evolution of $|\omega|$ during the late stages for a simulation in the weak vortex regime $I_P=0.25$ with $Re_\Gamma=1000$ (top row) and $Re_\Gamma=3000$ (bottom row). }
    \label{fig:ww-ip0p25-late}
\end{figure}

\begin{figure}
    \centering
    \includegraphics[width=0.40\textwidth]{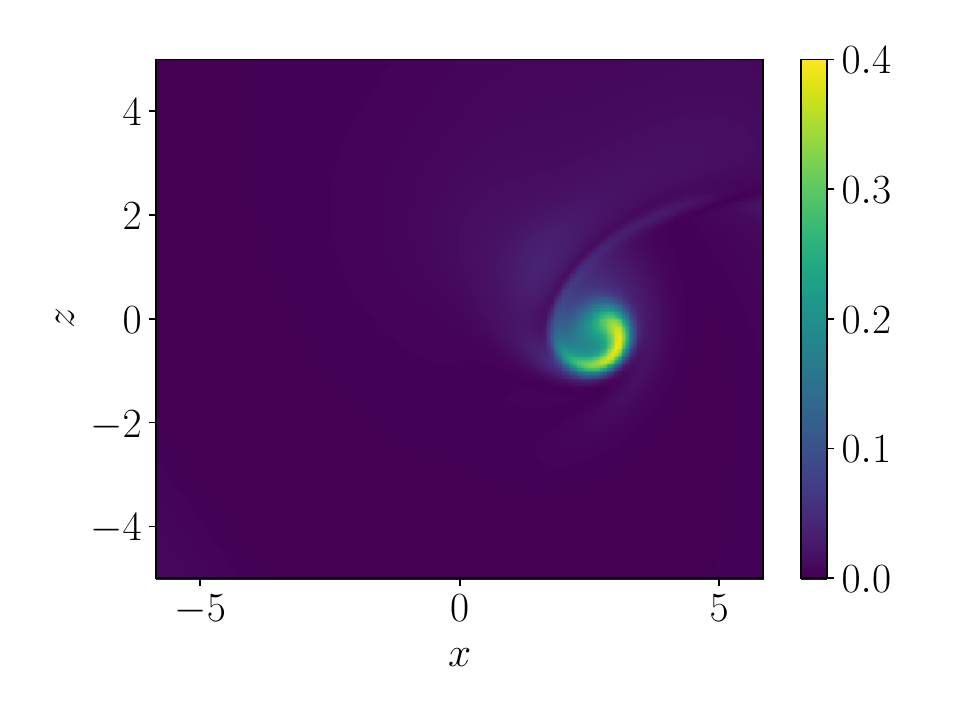}
    \includegraphics[width=0.40\textwidth]{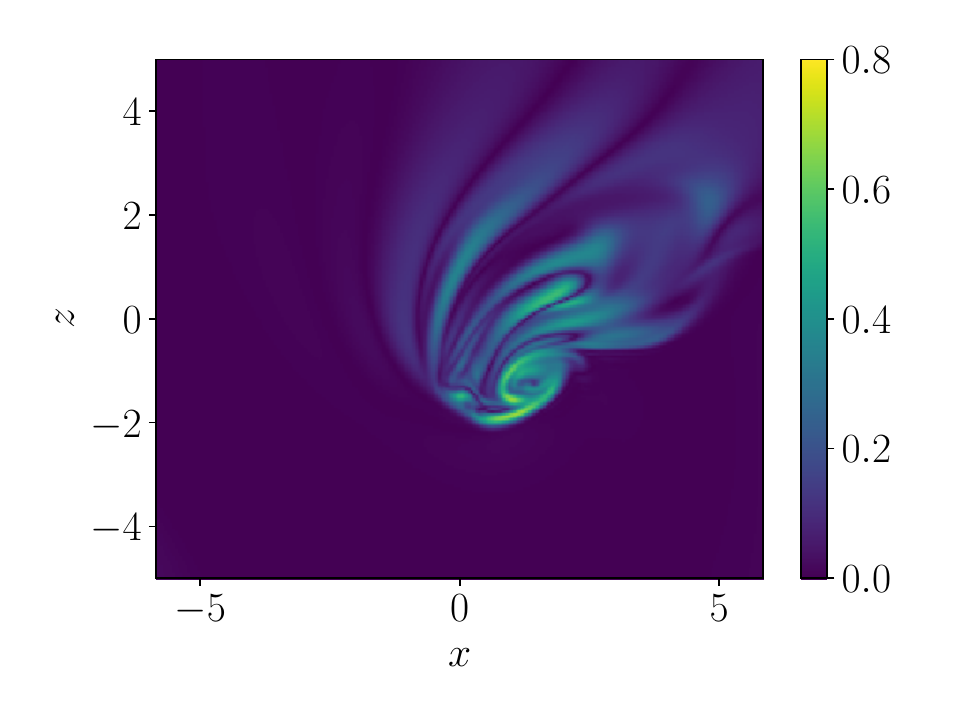}
    \caption{Vorticity magnitude $|\omega|$ at the $y=0$ mid-plane evolution and $t\Gamma/D^2=400$ for a simulation in the weak vortex regime at $I_P=0.25$ with $Re_\Gamma=1000$ (left) and with $Re_\Gamma=3000$ (right). }
    \label{fig:ww-ycut-ip0p25-late}
\end{figure}

The flow phenomenology after the wire has completely transversed the vortex is shown in full three dimensional visualizations in Figure \ref{fig:ww-ip0p25-late}, and in a selected cut in the $y=0$ plane in Figure \ref{fig:ww-ycut-ip0p25-late}. As the wire passes through the vortex, it leaves behind a wake. The waves reach the limit of the computational domain around $t\Gamma/D^2=400$. The primary vortex reconstitutes itself, even if vortex loops are seen to detach from it, especially at high Reynolds number. The resulting distribution of vorticity is different from the original Gaussian, as it is now concentrated in a thin sheet and displaced to the right by a few vortex cores. Despite this, a comparison of the initial and final circulation of the vortex for $Re_\Gamma=1000$ reveals only a $5\%$ loss throughout the entire process, while at $Re_\Gamma=3000$ the core has lost $15\%$ of its original circulation, very likely to the small scale structures which arise and dissipate. This showcases the vortex's ability to withstand the interaction significantly better than the case at $I_P=0.05$, which suffered a $75\%$ loss of its original circulation at $Re_\Gamma=1000$, and was not significantly detectable at $Re_\Gamma=3000$. 

\begin{figure}
     \centering
     \includegraphics[width=0.45\textwidth]{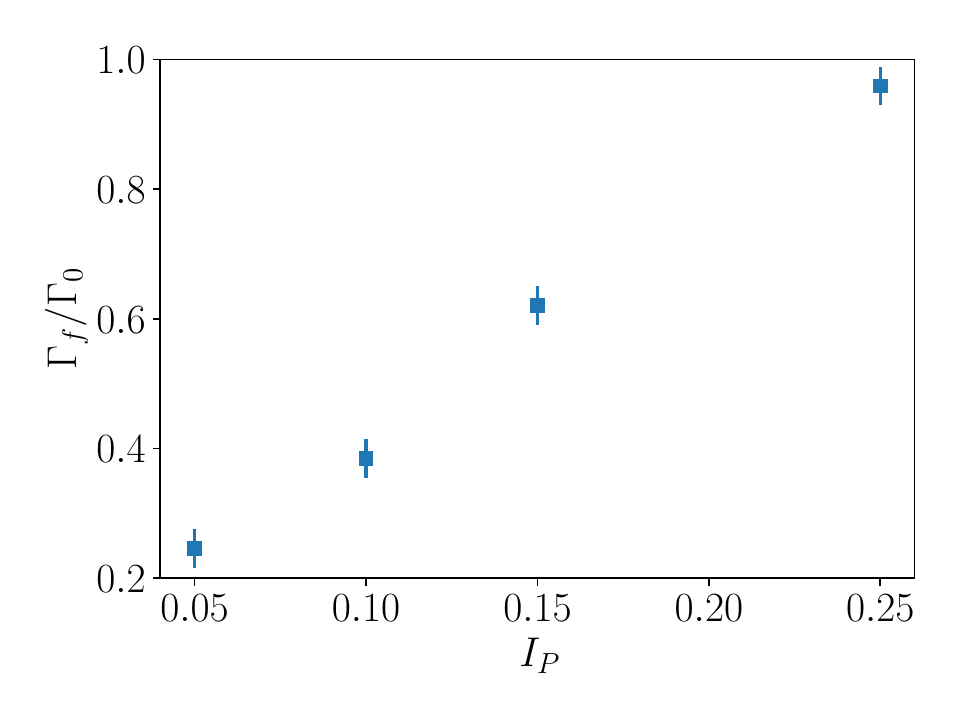}
     \caption{Loss of circulation during the interaction quantified as the ratio between final and initial circulations as a function of $I_P$ for the four cases simulated at $Re_\Gamma=1000$. }
     \label{fig:gammaf-re1000}
\end{figure}

The loss of circulation after the interaction has taken place ($\Gamma_f$) can be quantified for all cases at $Re_\Gamma=1000$. Figure \ref{fig:gammaf-re1000} shows $\Gamma_f/\Gamma_0$, $\Gamma_0$ being the original circulation of the vortex. The amount of circulation in the vortex by the end of the interaction monotonically increases with $I_P$ in the range studied. We note that this trend cannot persist, as the circulation must saturate at $\Gamma_f\approx\Gamma_0$ at higher values of $I_P$, as it is not expected that the wire imparts additional circulation to the vortex.

From these findings, it becomes evident that the primary control parameter in BVI is the impact parameter $I_P$. Varying the Reynolds number results in vorticity intensification and amplified distortion along the primary vortex, driven either by secondary vorticity or wave propagation. Also, at the later stages finer flow structures and turbulent clouds appear at higher Reynolds number, due to the increased instability of the vortices. However, these alterations do not significantly impact the fundamental collision features, at least up to $Re_\Gamma=3000$. Furthermore, our simulations successfully reproduce and provide reasoning for most experimental outcomes, demonstrating that even in the absence of turbulence, the phenomenology of weak and strong BVI can be faithfully replicated.


\section{Forces on the wire}
\label{sec:force}

\subsection{Initial Remarks}

This section discusses an area which is much less analyzed in BVI: measurement of the force on the cylinder. However, an adequate definition of the force on is not simple. Two things have to be taken into account: first, the force coming from BVI is a deviation from a constant drag which resists the movement of the cylinder at a constant velocity in a quiescent flow. Secondly, this deviation will be localized to the point of interaction, and if the forces for a quasi-infinite cylinder or blade undergoing BVI are added up, they would be approximately the same for those where there is no BVI, as the constant drag would overwhelm the forces from BVI.

\begin{figure}
    \centering
    \includegraphics[width=0.32\textwidth]{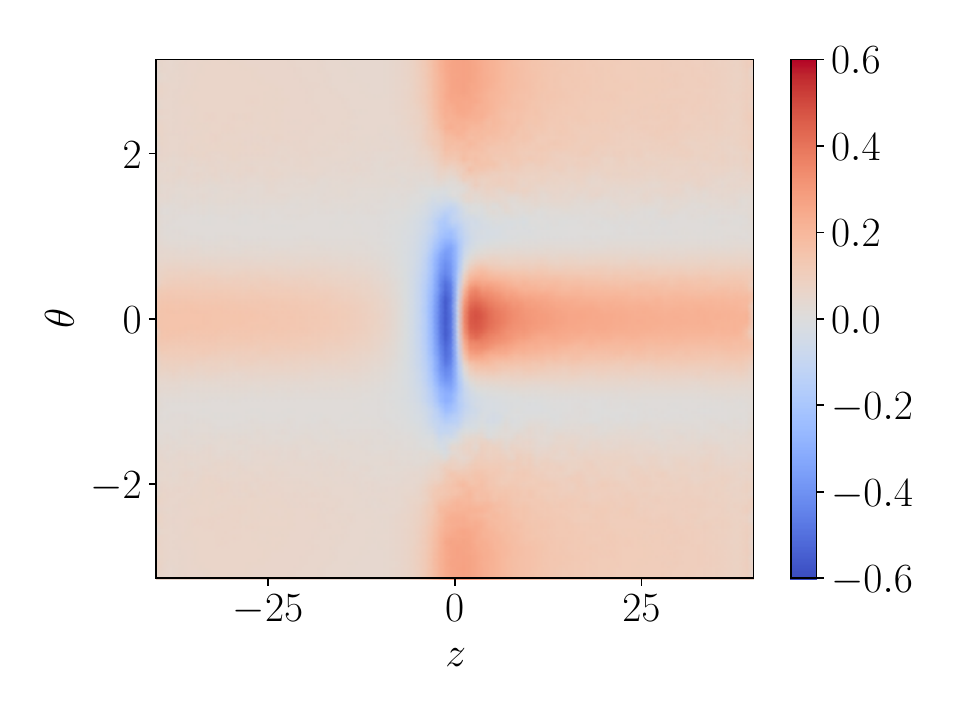}
    \includegraphics[width=0.32\textwidth]{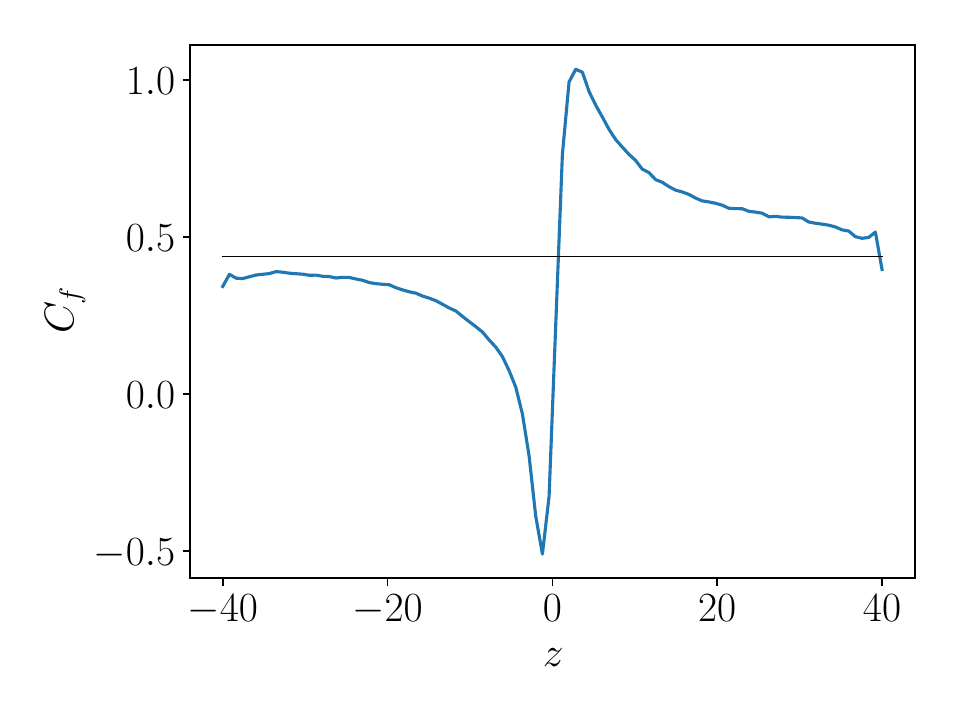}
    \includegraphics[width=0.32\textwidth]{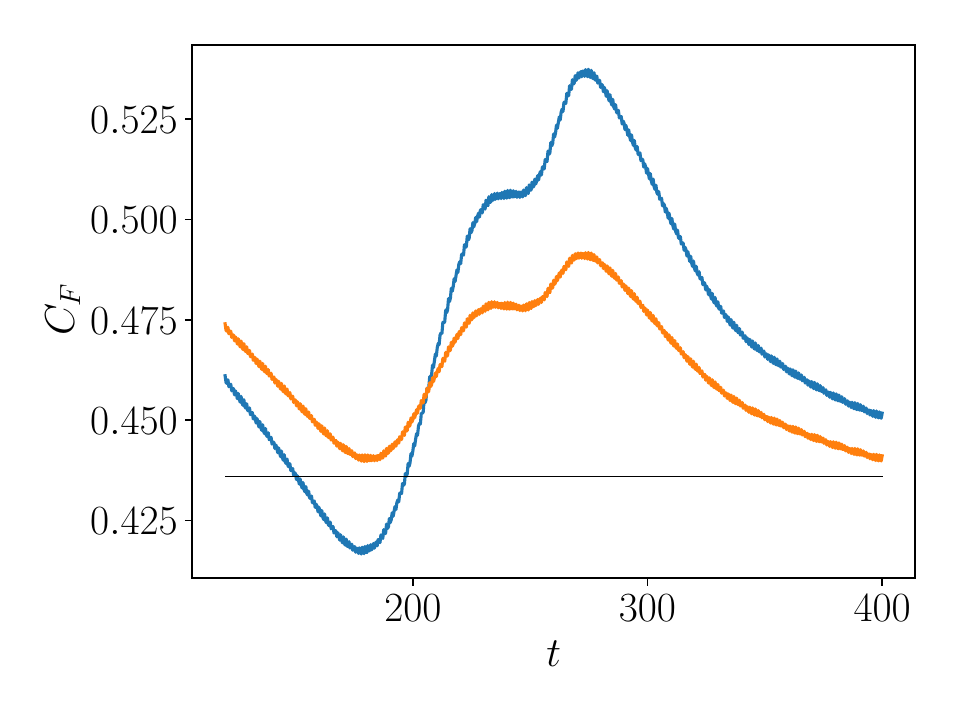}
    \caption{Left: Pressure coefficient along the cylinder for $Re_\Gamma=1000$, $I_P=0.25$ at time $t\Gamma/D^2=160$. Middle: Force coefficient along the cylinder axis for the same case. Left: Evolution of drag coefficient with time for the full cylinder (orange) and half the cylinder (blue).}
    \label{fig:force-bvi-intro}
\end{figure}

To elaborate on this, the left panel of Figure \ref{fig:force-bvi-intro} shows the force coefficient distribution $C_{f,x}=f_x/(\frac{1}{2}\rho [\Gamma/\sigma]^2)$ on the cylinder for the weak vortex case ($I_P=0.25$, $Re_\Gamma=1000$) before the cylinder has reached the vortex core. We note that this representation is highly distorted as the cylinder is very long. However, it is useful to note a few things. First, the distribution is symmetric around $\theta=0$, which corresponds to the leading edge of the cylinder. This is a natural product of the symmetry of the system. Second, while the region of strongest interaction is closely localized to the impact point, as could be expected, the return to asymptotic values is rather slow. Finally, a rather small region of the cylinder sees an overall force in the direction of motion of motion of the cylinder ($C_{f,x}>0$). This is due to the fact that locally the flow direction can be reversed when the velocity induced due to the vortex is larger than the cylinder motion.  
 
These results are emphasized in the middle panel of Figure \ref{fig:force-bvi-intro}, which shows the force on the cylinder along its axis $C_f(z)$. The baseline case where no vortex is present is shown as a solid line. The forces can be seen to be somewhat symmetrical around $z=0$, i.e.~the collision point. The overall force again shows positive values, coincident with regions where the velocity the cylinder sees is reversed. Finally, a long tail decay is observed: the forces do not return to the baseline case even at the edges of the cylinder. We note that the simulations yield qualitatively similar curves to those presented in the simplified models of Ref.~\cite{affes1993model_part_1}, with the presence of a suction peaks and regions of increased pressure. However, comparison is difficult due to the simplifying assumptions made of the flow and the different parameter regimes investigated.

As a final step, the average drag coefficient can be calculated. If the entire extent of the cylinder is used, a value of $C_F=0.449$ is obtained. If the cylinder half closest to the vortex is used, and the two quarters close to the edges are discarded, a value of $C_F=0.428$ is obtained. While both values are almost indistinguishable from the baseline value of $C_F=0.436$, one deviates upwards and the other downwards. These problems of adequately calculating $C_F$ are showcased in the right panel of Figure \ref{fig:force-bvi-intro}, which shows the time evolution of $C_F$ calculated using the full cylinder and half the cylinder. Both values show similar trends of increasing and decreasing during stages of the collision. However, they can be either above or below the baseline value depending on the cylinder segment considered for calculation. In any case, the deviations are mild, and only reach a maximum of $~20\%$ drag increase for the half-cylinder calculation. We also note that some small oscillations are present in the force calculation, which have a frequency close to the natural frequency of the cylinder $D/V$. 

Due to the problems of adequately defining $C_F$, we choose not to represent it for other cases and will only show the evolution of $C_f(z)$ for a few select cases below.

\subsection{Results}

\begin{figure}
    \centering
    \includegraphics[width=0.45\textwidth]{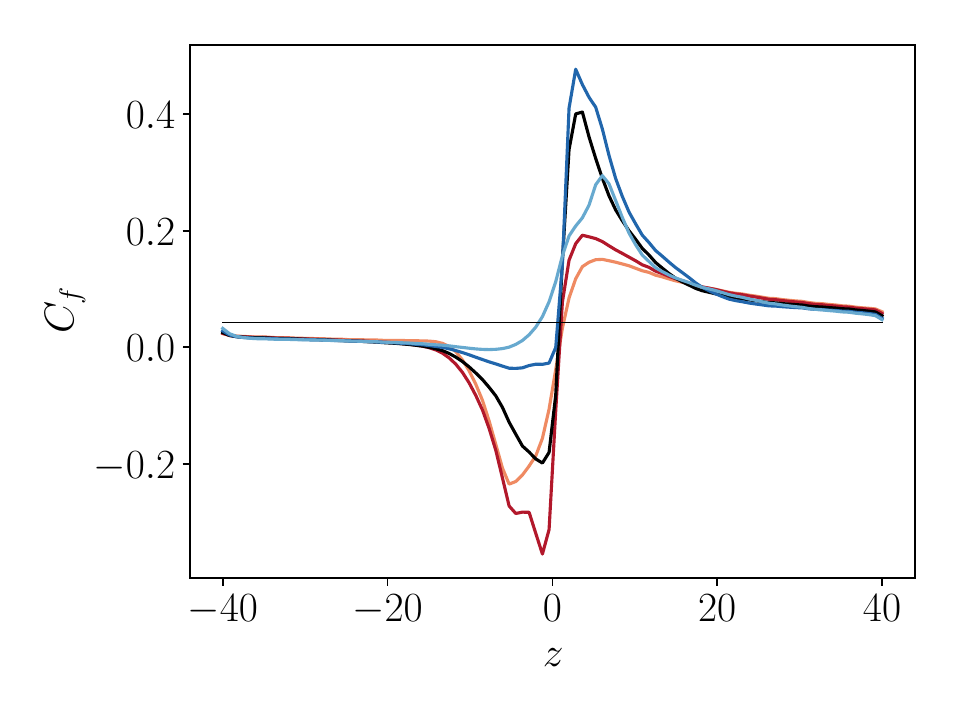}
    \includegraphics[width=0.45\textwidth]{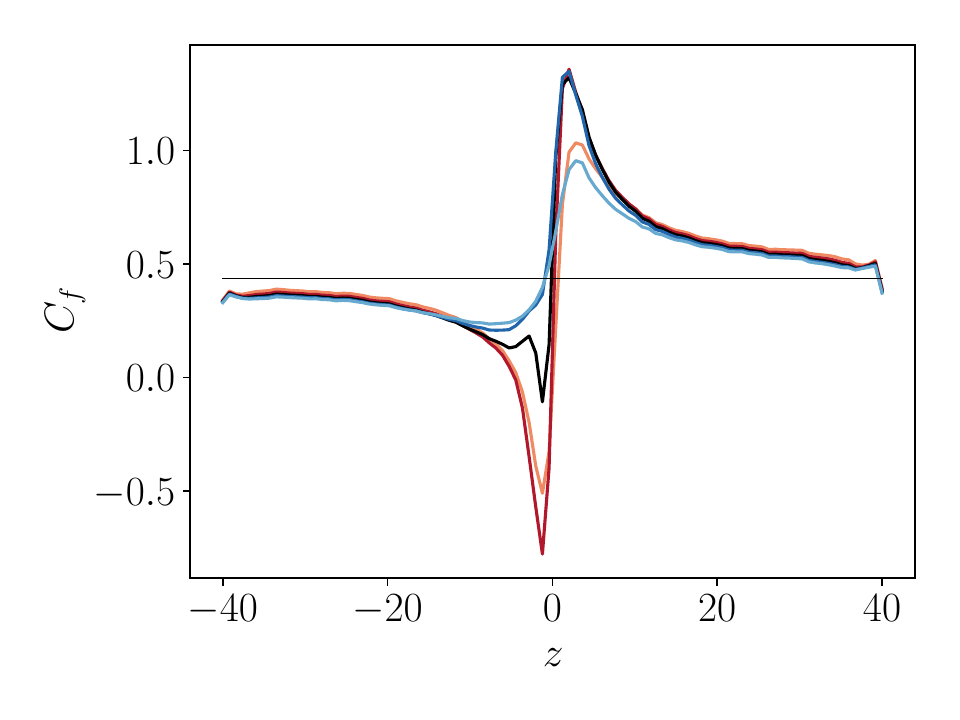}
    \includegraphics[width=0.45\textwidth]{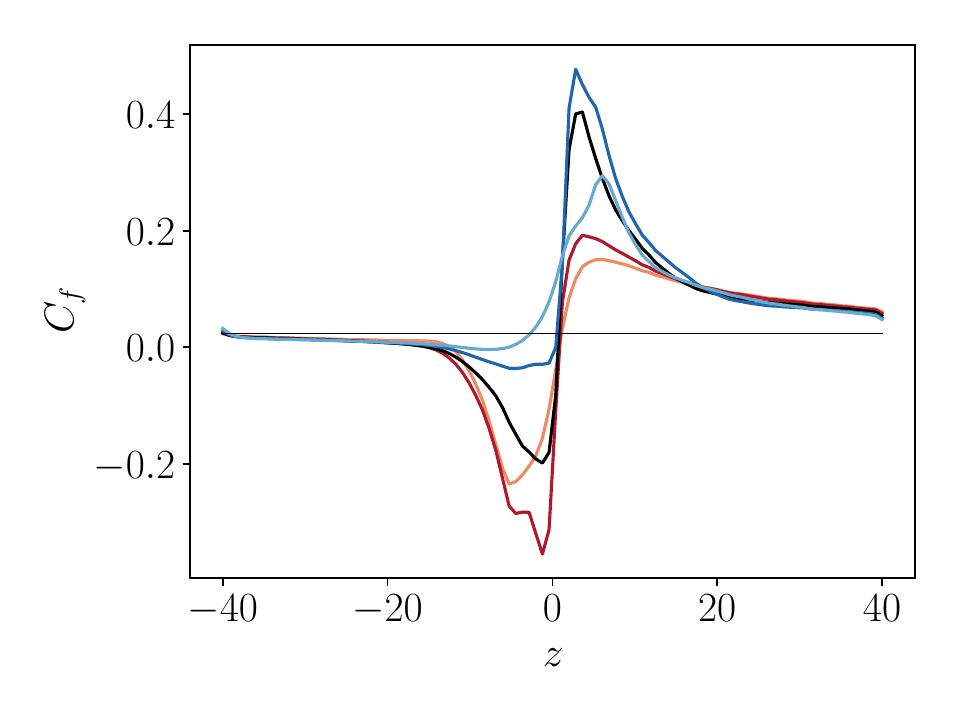}
    \includegraphics[width=0.45\textwidth]{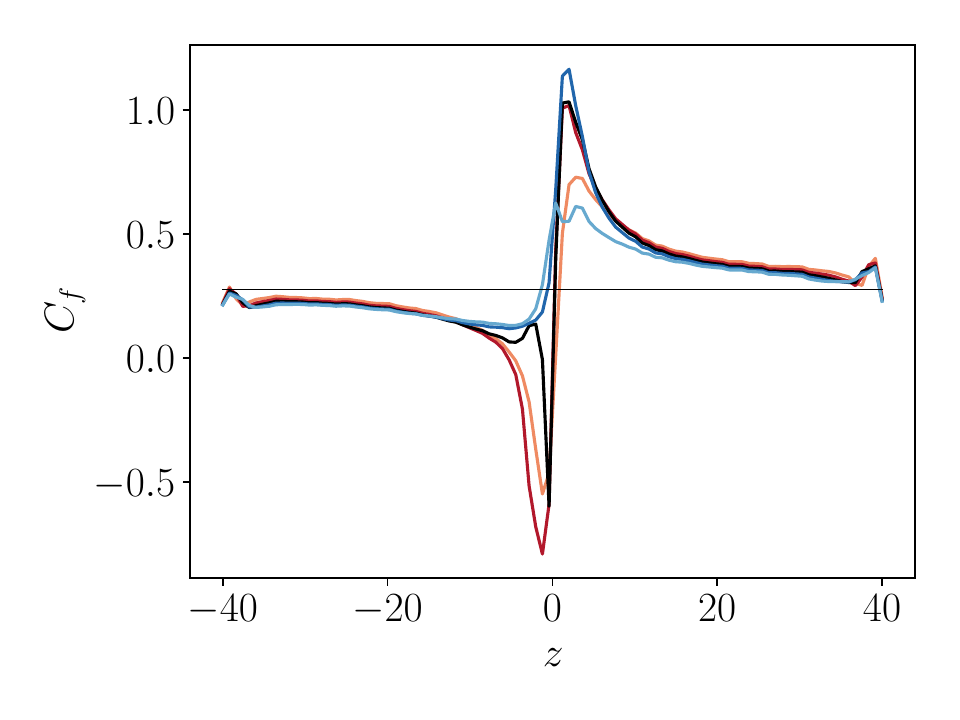}
    \caption{Force coefficient along the cylinder axis at several time instances. Left column is cases at $I_P=0.05$, while right column is cases at $I_P=0.25$. Top row is at $Re_\Gamma=1000$, while bottom row is at $Re_\Gamma=3000$. Symbols: left column ($I_P=0.05$), light to dark: $t\Gamma/D^2=700$ (light red), $1100$ (dark red), $1500$ (black), $1900$ (dark blue) and $2300$ (blue). Right column ($I_P=0.25$): $t\Gamma/D^2=160$ (light red), $200$ (dark red), $240$ (black), $280$ (dark blue) and $320$ (light blue). The horizontal thin line represents the baseline value. }
    \label{fig:force-bvi-results}
\end{figure}

In this section, $C_f(z)$ is calculated for a few selected time instants for four characteristic cases. These are two cases for each of the strong ($I_P=0.05$) and weak ($I_P=0.25$) vortex regimes, at the two Reynolds numbers simulated. These are shown in Figure \ref{fig:force-bvi-results}. 

The discussion starts with the case with $I_P=0.05$ and $Re_\Gamma=1000$. The curves are relatively similar to those shown in the previous subsection. The curves deviate from the baseline value, and they are roughly antisymmetrical around the collision point $z=0$. The deviations increase as the cylinder approaches the vortex, but they reach a maximum at distinct times: the peak force inversion ($C_f>0$) happens earlier than the peak suction ($C_f<0$). In the same manner, the values in the force inversion side return to the baseline values earlier than in the suction side. This is probably due to the fact that the cylinder boundary layer is being lifted up earlier from the force inversion side, so the interaction happens first on this half of the cylinder. Finally, the deviations from the baseline value are much more significant in this case: after all, the vortex is strong. This means that it will create forces on the cylinder which are much larger than those originating solely from the ordinary drag force.

Turning to the case with $I_P=0.25$ and $Re_\Gamma=1000$, the behaviour of the force has some similarities but also some differences when compared to the earlier case. The main difference is that the forces in the regions of large positive $C_f$ fluctuate much less- it attains a maximum value and remains close to that maximum value for most of the interaction. This suggests a probable absence of significant boundary layer liftoff. On the cylinder half experiencing force reversal, temporal variations are more pronounced yet localized due to the weaker vortex. In essence, little more can be extrapolated from these observations.

Finally, the bottom panels of Figure \ref{fig:force-bvi-results} show the Reynolds number dependence of the force, which can be contrasted to the top row. There is a relatively weak dependence of behavior on Reynolds number. The primary difference lies in the smaller baseline values of $C_F$, yet both curves exhibit analogous behaviour. Close to the impact point, spatial oscillations in $C_f(z)$ become more discernible, likely owing to the influence of smaller and more intense vortical structures on the cylinder.

In essence, our investigation into the forces on the cylinder reveals that while locally, BVI can generate substantial forces, especially within the strong vortex regime, these forces tend to average out across the cylinder halves facing either positive or negative vortex velocities. Consequently, the average force remains relatively consistent.

\section{Summary and conclusions}
\label{sec:conc}

Results from direct numerical simulations of normal body-vortex interaction using a thin cylinder have been presented. The focus was on the transition and the differences in the weak and strong vortex regimes. Three findings were reported:

\begin{enumerate}
    \item The simulations confirm that the transition between weak and strong vortex regimes is a smooth transition controlled by $I_P$ and primarily driven by the capacity of the cylinder's boundary layer to detach and wrap around the primary vortex before impact takes place. 
    \item In the strong vortex regime, this detached, or secondary vorticity propagates along the vortex's axis, interacting with it and heavily debilitating it, with the result that by the end of the interaction the vortex has almost dissipated. 
    \item In the weak vortex regime, the cylinder only interacts with the vortex once the body reaches the tube. It then deforms it, propagating waves along the axis. The behavior of the waves was significantly different to those observed when axial flow is present \cite{krishnamoorthy1998three}. 
\end{enumerate}

The Reynolds number dependence of the flow was also investigated by simulating two cases at $Re_\Gamma=3000$. The effect of $Re_\Gamma$ on the system was reflected in two aspects: at higher $Re_\Gamma$ the boundary layers are thinner and have more vorticity, and vortices are more unstable and prone to breaking down. Three main results were found:

\begin{enumerate}
   \item In the strong vortex regime, qualitative differences in the flow can be seen before the impact of the cylinder. The secondary vorticity, which is now concentrated in smaller and more intense structures, has a stronger interaction with the primary tube leading to more deformations in the neck area. These differences are absent from the weak vortex regime due to the reduced importance of secondary vorticity originating from the boundary layer in the early stages of the interaction.
   \item In the weak vortex regime, the stretching and necking of the primary vortex due to the cylinder forcing oppositely signed vorticity close to it is enhanced. However, this does not result in significantly different outcomes.
   \item After the interaction, the vortex remnants tend to break down rather than recompose again into a primary tube. The simulations indicate that even in the absence of background perturbations, the flow left behind the cylinder will be turbulent. 
\end{enumerate}

The results presented in this manuscript qualitatively reproduce existing experimental data \cite{marshall1997instantaneous, krishnamoorthy1998three} when the presence of axial flow is not important.  The simulations also confirm that the adequate modelling of the boundary layer is necessary to capture the intricacies of the strong vortex regime. 

The distribution of pressure forces on the cylinder was also reported. The results show local deviations from the baseline value (considered as the one where no vortex is present), which are concentrated around the impact point. The deviations are large, to the point that the force can locally reverse direction, with the cylinder feeling a force in the direction of the motion. However, in general the forces average out on both sides of the interaction, and the integrated force coefficients are very similar to the baseline value.  

As the introduction stated, the starting point of the study was on ``cutting'' vortex lines. However, these simulations showed that the phenomenology is much more complicated, especially for the strong vortex case, and that the interaction deviates from what one could expect from ideal vortex models. While the scope of the manuscript has been restricted to normal BVI, with minor modifications the code could be extended to simulating parallel and grazing BVI, and regimes in between. Another possible extension of this work is the question of what happens when a vortex ring- and not a vortex tube is cut. Very few studies of cutting a vortex ring with a wire have been made \cite{naitoh1995vortex}. Novel structures such as ``elephant ears'' have been reported, which have not been related to the wider literature on BVI. 

\begin{acknowledgments}
We thank the Research Computing Data Core (RCDC) at the University of Houston for providing computing resources. 
\end{acknowledgments}

\appendix 

\section{Grid dependence studies}
\label{app:grid}

To assess the adequateness of the grid, two additional simulations of the case with $I_P=0.25$ and $Re_\Gamma=3000$ were conducted with coarse and fine grids. These parameters were chosen as this simulation case was the most unfavourable as the cylinder Reynolds number was highest ($Re_c=113$), so it provides an upper limit for the problems with the grid. The coarse grid is obtain by halving the resolution in each spatial direction, i.e.~$312\times312\times936$ points, while the fine grid is obtained by increasing the resolution by a factor $1.5\times$ in every direction, i.e.~$936\times936\times2808$. 

\begin{figure}
    \centering
    \includegraphics[width=0.32\textwidth]{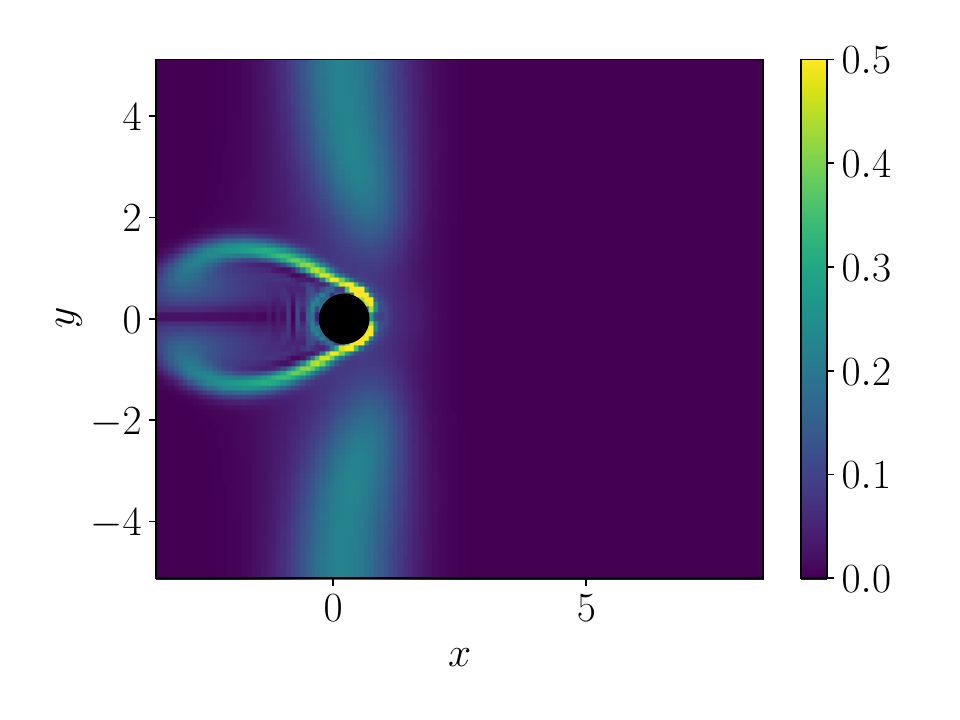}
    \includegraphics[width=0.32\textwidth]{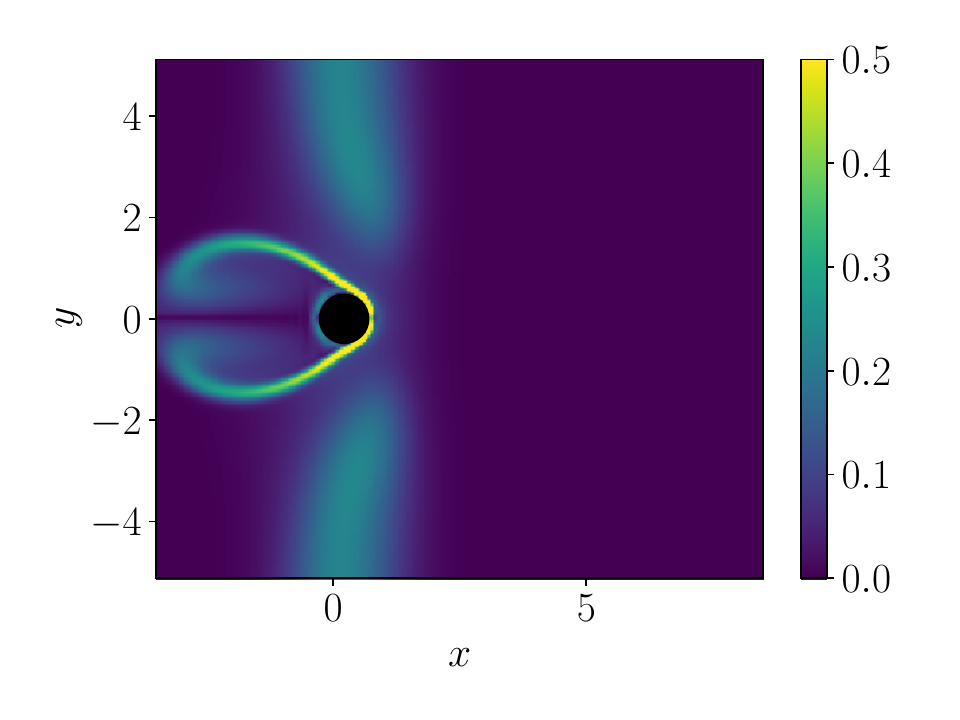}
    \includegraphics[width=0.32\textwidth]{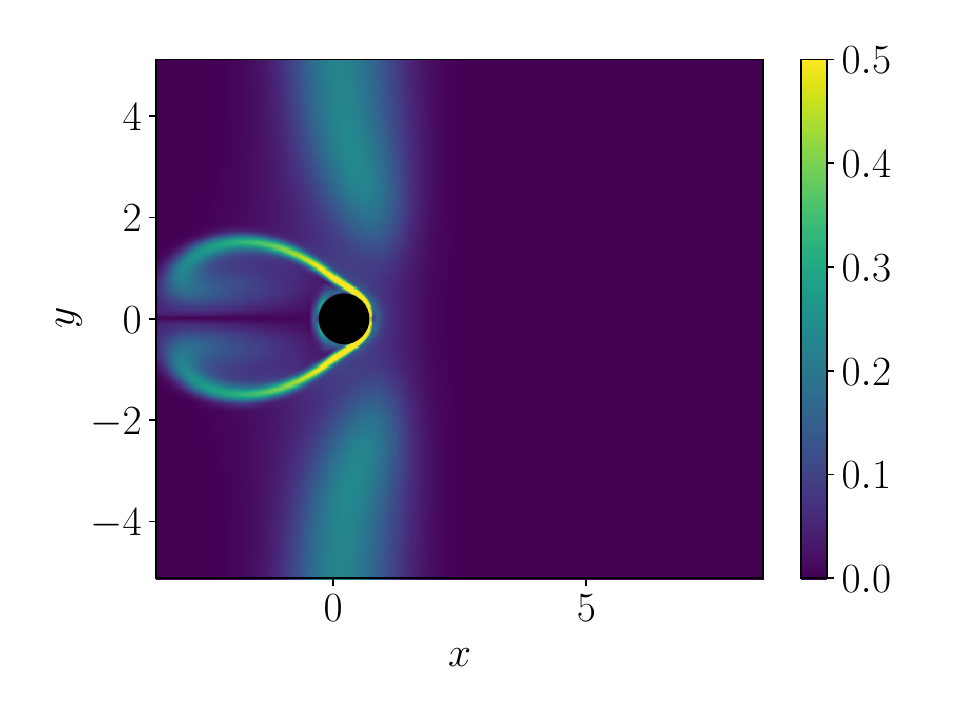}\\
    \includegraphics[width=0.32\textwidth]{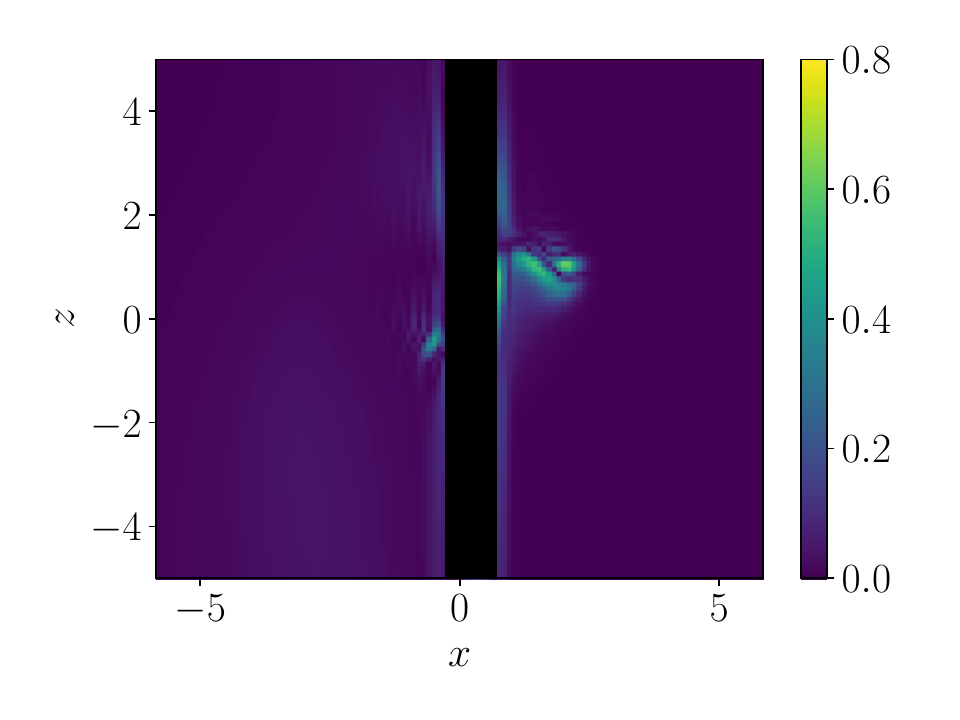}
    \includegraphics[width=0.32\textwidth]{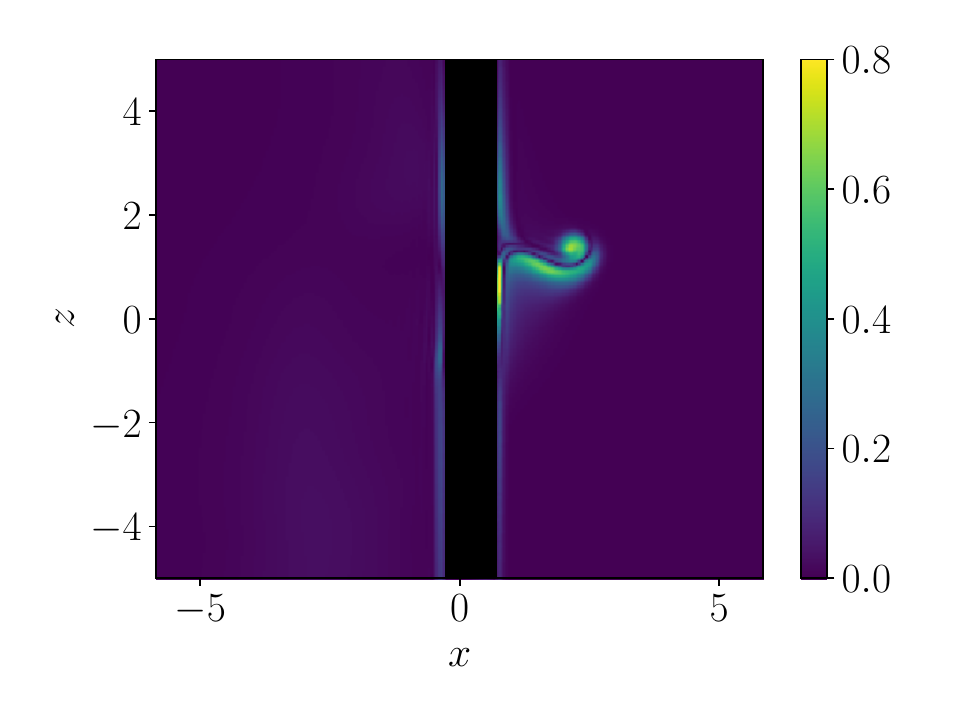}
    \includegraphics[width=0.32\textwidth]{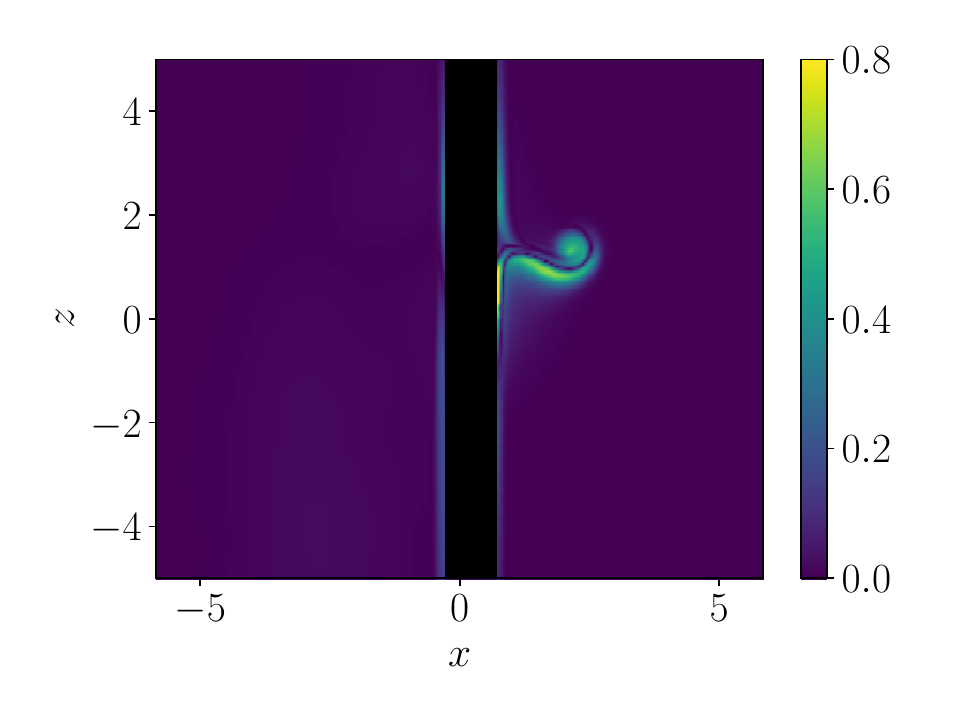}\\    \includegraphics[width=0.32\textwidth]{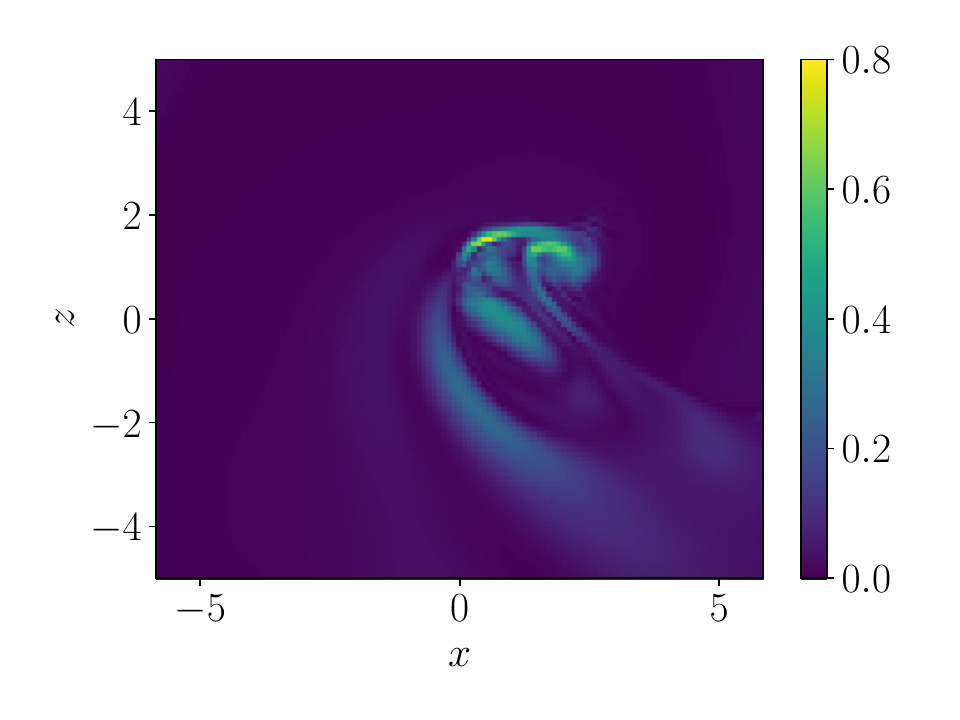}
    \includegraphics[width=0.32\textwidth]{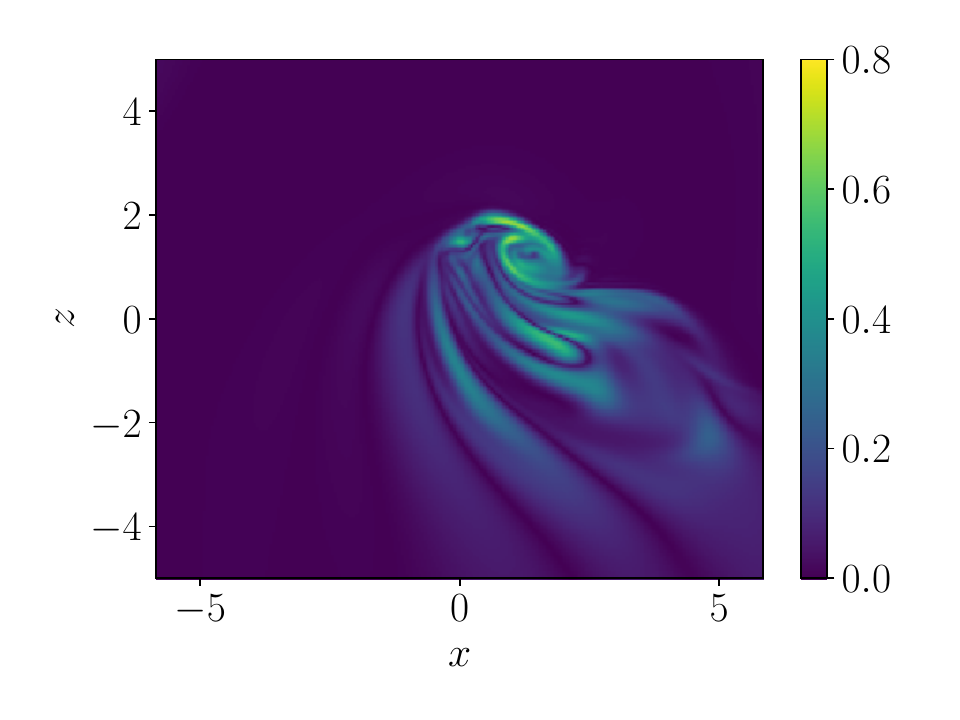}
    \includegraphics[width=0.32\textwidth]{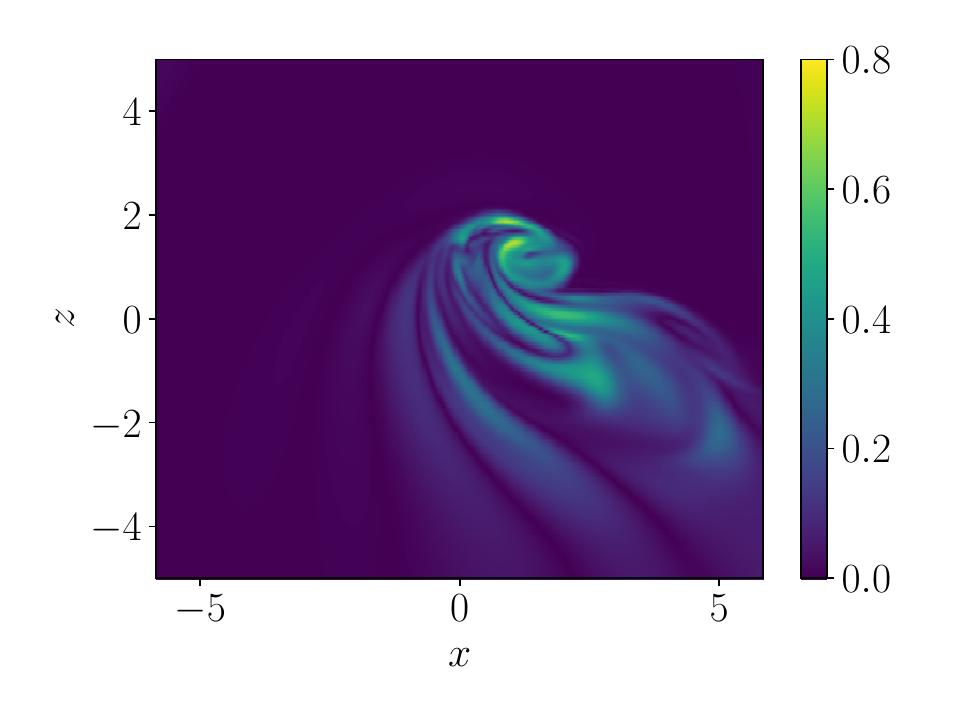}
    \caption{Vorticity magnitude for the $I_P=0.25$ and $Re_\Gamma=3000$ case at $t\Gamma^2/D=220$ and $z=0$ (top), at $t\Gamma^2/D=220$ and $y=0$ (middle), at $t\Gamma^2/D=220$ and $y=0$ (bottom). The left, middle and right columns show the coarse, medium and fine resolutions respectively. }
    \label{fig:ww-cut-rescheck}
\end{figure}

Figure \ref{fig:ww-cut-rescheck} shows the vorticity magnitude for the three grids alongside two plane cuts. The first two rows show the variable at $t\Gamma^2/D=220$: this time instant was chosen because it is close to peak dissipation, the tube has been stretched and displaced, and vorticity has intensified which makes it challenging to resolve. Cuts are shown for both $z=0$, i.e.~across the wire, $y=0$, i.e.~across the vortex core. Clear signatures of underresolution can be seen for the coarse mesh: there are marks of numerical dispersion behind the cylinder in the top-left panel, as well as deformations in the vortex core not seen for the medium and fine grids. While some differences to appear between the medium and fine grids, they are not as significant. This can be quantified by measuring the total circulation in the vortex cores shown in the right column: the coarse grid has a circulation which is $25\%$ lower than the medium grid, while the fine grid has a circulation which is $2\%$ higher than the medium grid. This confirms our assessment that the medium grid can be considered is adequate to within $2\%$, but that further coarsening gives results which are inaccurate due to numerical dispersion.

Finally, the bottom row shows the vorticity magnitude at $t\Gamma^2/D=400$ through the $y=0$ plane. This is after the interaction has finished, so the differences between cases should be maximum. Once again, the medium and fine grids show similarities, while the coarse grid shows a picture that is somewhat similar to the other two grids, yet with vorticity concentrated in different points.

\begin{figure}
    \centering
    \includegraphics[width=0.45\textwidth]{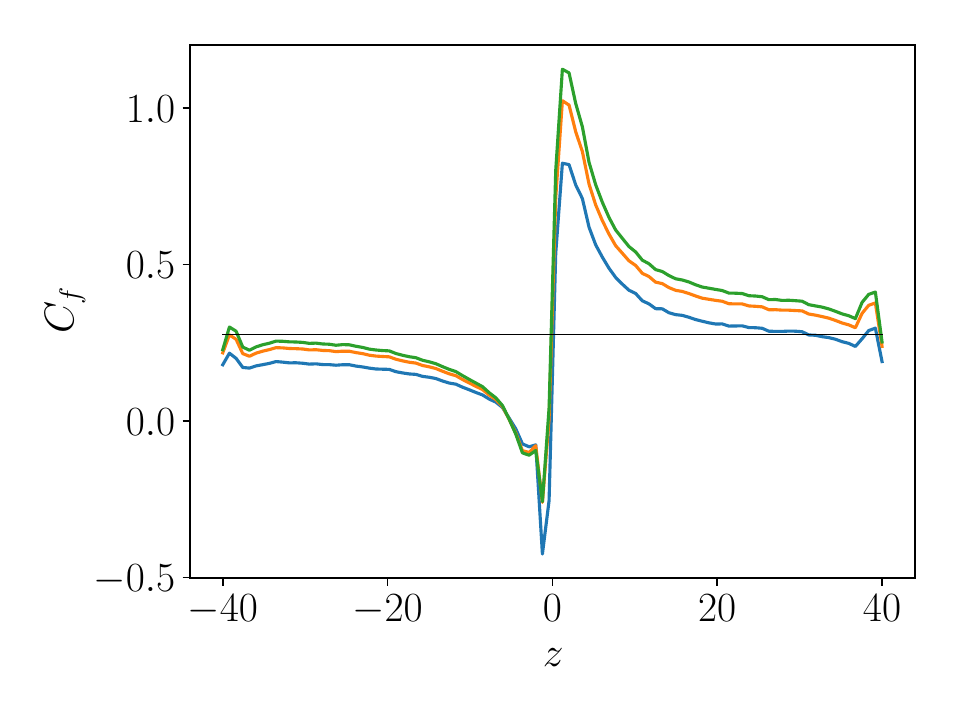}
    \caption{Force coefficient at $t\Gamma^2/D=220$ for the $I_P=0.25$ and $Re_\Gamma=3000$ case for the three grids. A horizontal line shows the baseline value when no vortex is present. Symbols: blue, coarse grid; orange, medium grid; green, fine grid. }
    \label{fig:fz-rescheck}
\end{figure}

A further check was conducted by assessing the variation in the force on the wire. Figure \ref{fig:fz-rescheck} shows $C_f(z)$ at the same time instant for the three meshes. Again, differences can be seen between the three cases, but the difference between the coarse grid and the medium and fine grids is much larger than the difference between medium and fine grids. The integrated force coefficients are $C_F=0.24$, $C_F=0.31$ and $C_F=0.33$ for the coarse, medium and fine grids respectively. This shows that the quantities on the wire are slightly more sensitive than the vorticity statistics. However, the curves all show similar phenomenology, with a strong suction peak, and a strong increase close to the vortex core. Hence, we do not expect the results from the medium mesh to miss any important physics, while the coarse mesh shows $C_f$ dropping below the baseline value, questioning its use. 

To summarize, the medium mesh shows small discrepancies when compared to a finer mesh, especially when measuring the force on the wire. However, the overall phenomenology is well captured, unlike what is seen for the coarser mesh, where the flow topology is changed. We also reiterate that this is the most unfavourable instance, and yet the medium grid is able to produce reasonable results with a much lower computational cost.

\clearpage

\bibliography{apssamp}

\begin{thebibliography}{42}%
\makeatletter
\providecommand \@ifxundefined [1]{%
 \@ifx{#1\undefined}
}%
\providecommand \@ifnum [1]{%
 \ifnum #1\expandafter \@firstoftwo
 \else \expandafter \@secondoftwo
 \fi
}%
\providecommand \@ifx [1]{%
 \ifx #1\expandafter \@firstoftwo
 \else \expandafter \@secondoftwo
 \fi
}%
\providecommand \natexlab [1]{#1}%
\providecommand \enquote  [1]{``#1''}%
\providecommand \bibnamefont  [1]{#1}%
\providecommand \bibfnamefont [1]{#1}%
\providecommand \citenamefont [1]{#1}%
\providecommand \href@noop [0]{\@secondoftwo}%
\providecommand \href [0]{\begingroup \@sanitize@url \@href}%
\providecommand \@href[1]{\@@startlink{#1}\@@href}%
\providecommand \@@href[1]{\endgroup#1\@@endlink}%
\providecommand \@sanitize@url [0]{\catcode `\\12\catcode `\$12\catcode
  `\&12\catcode `\#12\catcode `\^12\catcode `\_12\catcode `\%12\relax}%
\providecommand \@@startlink[1]{}%
\providecommand \@@endlink[0]{}%
\providecommand \url  [0]{\begingroup\@sanitize@url \@url }%
\providecommand \@url [1]{\endgroup\@href {#1}{\urlprefix }}%
\providecommand \urlprefix  [0]{URL }%
\providecommand \Eprint [0]{\href }%
\providecommand \doibase [0]{https://doi.org/}%
\providecommand \selectlanguage [0]{\@gobble}%
\providecommand \bibinfo  [0]{\@secondoftwo}%
\providecommand \bibfield  [0]{\@secondoftwo}%
\providecommand \translation [1]{[#1]}%
\providecommand \BibitemOpen [0]{}%
\providecommand \bibitemStop [0]{}%
\providecommand \bibitemNoStop [0]{.\EOS\space}%
\providecommand \EOS [0]{\spacefactor3000\relax}%
\providecommand \BibitemShut  [1]{\csname bibitem#1\endcsname}%
\let\auto@bib@innerbib\@empty
\bibitem [{\citenamefont {Saffman}(1995)}]{saffman1995vortex}%
  \BibitemOpen
  \bibfield  {author} {\bibinfo {author} {\bibfnamefont {P.~G.}\ \bibnamefont
  {Saffman}},\ }\href@noop {} {\emph {\bibinfo {title} {Vortex Dynamics}}}\
  (\bibinfo  {publisher} {Cambridge University Press},\ \bibinfo {year}
  {1995})\BibitemShut {NoStop}%
\bibitem [{\citenamefont {Kida}\ and\ \citenamefont
  {Takaoka}(1994)}]{kida1994vortex}%
  \BibitemOpen
  \bibfield  {author} {\bibinfo {author} {\bibfnamefont {S.}~\bibnamefont
  {Kida}}\ and\ \bibinfo {author} {\bibfnamefont {M.}~\bibnamefont {Takaoka}},\
  }\bibfield  {title} {\bibinfo {title} {Vortex reconnection},\ }\href@noop {}
  {\bibfield  {journal} {\bibinfo  {journal} {Annual Review of Fluid
  Mechanics}\ }\textbf {\bibinfo {volume} {26}},\ \bibinfo {pages} {169}
  (\bibinfo {year} {1994})}\BibitemShut {NoStop}%
\bibitem [{\citenamefont {Marshall}\ and\ \citenamefont
  {Yalamanchili}(1994)}]{marshall1994vortex}%
  \BibitemOpen
  \bibfield  {author} {\bibinfo {author} {\bibfnamefont {J.}~\bibnamefont
  {Marshall}}\ and\ \bibinfo {author} {\bibfnamefont {R.}~\bibnamefont
  {Yalamanchili}},\ }\bibfield  {title} {\bibinfo {title} {Vortex cutting by a
  blade. {{{II-C}}}omputations of vortex response},\ }\href@noop {} {\bibfield
  {journal} {\bibinfo  {journal} {AIAA journal}\ }\textbf {\bibinfo {volume}
  {32}},\ \bibinfo {pages} {1428} (\bibinfo {year} {1994})}\BibitemShut
  {NoStop}%
\bibitem [{\citenamefont {Naitoh}\ \emph {et~al.}(1995)\citenamefont {Naitoh},
  \citenamefont {Sun},\ and\ \citenamefont {Yamada}}]{naitoh1995vortex}%
  \BibitemOpen
  \bibfield  {author} {\bibinfo {author} {\bibfnamefont {T.}~\bibnamefont
  {Naitoh}}, \bibinfo {author} {\bibfnamefont {B.}~\bibnamefont {Sun}},\ and\
  \bibinfo {author} {\bibfnamefont {H.}~\bibnamefont {Yamada}},\ }\bibfield
  {title} {\bibinfo {title} {A vortex ring travelling across a thin circular
  cylinder},\ }\href@noop {} {\bibfield  {journal} {\bibinfo  {journal} {Fluid
  Dynamics Research}\ }\textbf {\bibinfo {volume} {15}},\ \bibinfo {pages} {43}
  (\bibinfo {year} {1995})}\BibitemShut {NoStop}%
\bibitem [{\citenamefont {Liu}\ and\ \citenamefont
  {Marshall}(2004)}]{liu2004blade}%
  \BibitemOpen
  \bibfield  {author} {\bibinfo {author} {\bibfnamefont {X.}~\bibnamefont
  {Liu}}\ and\ \bibinfo {author} {\bibfnamefont {J.}~\bibnamefont {Marshall}},\
  }\bibfield  {title} {\bibinfo {title} {Blade penetration into a vortex core
  with and without axial core flow},\ }\href@noop {} {\bibfield  {journal}
  {\bibinfo  {journal} {Journal of Fluid Mechanics}\ }\textbf {\bibinfo
  {volume} {519}},\ \bibinfo {pages} {81} (\bibinfo {year} {2004})}\BibitemShut
  {NoStop}%
\bibitem [{\citenamefont {Rockwell}(1998)}]{rockwell1998vortex}%
  \BibitemOpen
  \bibfield  {author} {\bibinfo {author} {\bibfnamefont {D.}~\bibnamefont
  {Rockwell}},\ }\bibfield  {title} {\bibinfo {title} {Vortex-body
  interactions},\ }\href@noop {} {\bibfield  {journal} {\bibinfo  {journal}
  {Annual Review of Fluid Mechanics}\ }\textbf {\bibinfo {volume} {30}},\
  \bibinfo {pages} {199} (\bibinfo {year} {1998})}\BibitemShut {NoStop}%
\bibitem [{\citenamefont {Palau}\ \emph {et~al.}(2008)\citenamefont {Palau},
  \citenamefont {Dinner}, \citenamefont {Durrell},\ and\ \citenamefont
  {Blamire}}]{palau2008vortex}%
  \BibitemOpen
  \bibfield  {author} {\bibinfo {author} {\bibfnamefont {A.}~\bibnamefont
  {Palau}}, \bibinfo {author} {\bibfnamefont {R.}~\bibnamefont {Dinner}},
  \bibinfo {author} {\bibfnamefont {J.~H.}\ \bibnamefont {Durrell}},\ and\
  \bibinfo {author} {\bibfnamefont {M.~G.}\ \bibnamefont {Blamire}},\
  }\bibfield  {title} {\bibinfo {title} {Vortex breaking and cutting in type
  {{{II}}} superconductors},\ }\href@noop {} {\bibfield  {journal} {\bibinfo
  {journal} {Physical Review Letters}\ }\textbf {\bibinfo {volume} {101}},\
  \bibinfo {pages} {097002} (\bibinfo {year} {2008})}\BibitemShut {NoStop}%
\bibitem [{\citenamefont {Glatz}\ \emph {et~al.}(2016)\citenamefont {Glatz},
  \citenamefont {Vlasko-Vlasov}, \citenamefont {Kwok},\ and\ \citenamefont
  {Crabtree}}]{glatz2016vortex}%
  \BibitemOpen
  \bibfield  {author} {\bibinfo {author} {\bibfnamefont {A.}~\bibnamefont
  {Glatz}}, \bibinfo {author} {\bibfnamefont {V.}~\bibnamefont
  {Vlasko-Vlasov}}, \bibinfo {author} {\bibfnamefont {W.}~\bibnamefont
  {Kwok}},\ and\ \bibinfo {author} {\bibfnamefont {G.}~\bibnamefont
  {Crabtree}},\ }\bibfield  {title} {\bibinfo {title} {Vortex cutting in
  superconductors},\ }\href@noop {} {\bibfield  {journal} {\bibinfo  {journal}
  {Physical Review B}\ }\textbf {\bibinfo {volume} {94}},\ \bibinfo {pages}
  {064505} (\bibinfo {year} {2016})}\BibitemShut {NoStop}%
\bibitem [{\citenamefont {Jameson}(1970)}]{jameson1970analysis}%
  \BibitemOpen
  \bibfield  {author} {\bibinfo {author} {\bibfnamefont {A.}~\bibnamefont
  {Jameson}},\ }\bibfield  {title} {\bibinfo {title} {The analysis of
  propeller-wing flow interaction},\ }\href@noop {} {\bibfield  {journal}
  {\bibinfo  {journal} {Analytic methods in aircraft aerodynamics, number SP}\
  }\textbf {\bibinfo {volume} {228}},\ \bibinfo {pages} {721} (\bibinfo {year}
  {1970})}\BibitemShut {NoStop}%
\bibitem [{\citenamefont {Witkowski}\ \emph {et~al.}(1989)\citenamefont
  {Witkowski}, \citenamefont {Lee},\ and\ \citenamefont
  {Sullivan}}]{witkowski1989aerodynamic}%
  \BibitemOpen
  \bibfield  {author} {\bibinfo {author} {\bibfnamefont {D.~P.}\ \bibnamefont
  {Witkowski}}, \bibinfo {author} {\bibfnamefont {A.~K.}\ \bibnamefont {Lee}},\
  and\ \bibinfo {author} {\bibfnamefont {J.~P.}\ \bibnamefont {Sullivan}},\
  }\bibfield  {title} {\bibinfo {title} {Aerodynamic interaction between
  propellers and wings},\ }\href@noop {} {\bibfield  {journal} {\bibinfo
  {journal} {Journal of Aircraft}\ }\textbf {\bibinfo {volume} {26}},\ \bibinfo
  {pages} {829} (\bibinfo {year} {1989})}\BibitemShut {NoStop}%
\bibitem [{\citenamefont {Leishman}\ and\ \citenamefont
  {Bagai}(1998)}]{leishman1998challenges}%
  \BibitemOpen
  \bibfield  {author} {\bibinfo {author} {\bibfnamefont {J.~G.}\ \bibnamefont
  {Leishman}}\ and\ \bibinfo {author} {\bibfnamefont {A.}~\bibnamefont
  {Bagai}},\ }\bibfield  {title} {\bibinfo {title} {Challenges in understanding
  the vortex dynamics of helicopter rotor wakes},\ }\href@noop {} {\bibfield
  {journal} {\bibinfo  {journal} {AIAA Journal}\ }\textbf {\bibinfo {volume}
  {36}},\ \bibinfo {pages} {1130} (\bibinfo {year} {1998})}\BibitemShut
  {NoStop}%
\bibitem [{\citenamefont {Yung}(2000)}]{yung2000rotor}%
  \BibitemOpen
  \bibfield  {author} {\bibinfo {author} {\bibfnamefont {H.~Y.}\ \bibnamefont
  {Yung}},\ }\bibfield  {title} {\bibinfo {title} {Rotor blade--vortex
  interaction noise},\ }\href@noop {} {\bibfield  {journal} {\bibinfo
  {journal} {Progress in Aerospace Sciences}\ }\textbf {\bibinfo {volume}
  {36}},\ \bibinfo {pages} {97} (\bibinfo {year} {2000})}\BibitemShut {NoStop}%
\bibitem [{\citenamefont {Amet}\ \emph {et~al.}(2009)\citenamefont {Amet},
  \citenamefont {Ma{\^\i}tre}, \citenamefont {Pellone},\ and\ \citenamefont
  {Achard}}]{amet20092d}%
  \BibitemOpen
  \bibfield  {author} {\bibinfo {author} {\bibfnamefont {E.}~\bibnamefont
  {Amet}}, \bibinfo {author} {\bibfnamefont {T.}~\bibnamefont {Ma{\^\i}tre}},
  \bibinfo {author} {\bibfnamefont {C.}~\bibnamefont {Pellone}},\ and\ \bibinfo
  {author} {\bibfnamefont {J.-L.}\ \bibnamefont {Achard}},\ }\bibfield  {title}
  {\bibinfo {title} {{{{2D N}}}umerical simulations of blade-vortex interaction
  in a {{{Darrieus}}} turbine},\ }\href@noop {} {\bibfield  {journal} {\bibinfo
   {journal} {Journal of Fluids Engineering}\ }\textbf {\bibinfo {volume}
  {131}} (\bibinfo {year} {2009})}\BibitemShut {NoStop}%
\bibitem [{\citenamefont {Wilder}\ and\ \citenamefont
  {Telionis}(1998)}]{wilder1998parallel}%
  \BibitemOpen
  \bibfield  {author} {\bibinfo {author} {\bibfnamefont {M.}~\bibnamefont
  {Wilder}}\ and\ \bibinfo {author} {\bibfnamefont {D.}~\bibnamefont
  {Telionis}},\ }\bibfield  {title} {\bibinfo {title} {Parallel blade--vortex
  interaction},\ }\href@noop {} {\bibfield  {journal} {\bibinfo  {journal}
  {Journal of Fluids and Structures}\ }\textbf {\bibinfo {volume} {12}},\
  \bibinfo {pages} {801} (\bibinfo {year} {1998})}\BibitemShut {NoStop}%
\bibitem [{\citenamefont {Ahmadi}(1986)}]{Ahmadi1986AnEI}%
  \BibitemOpen
  \bibfield  {author} {\bibinfo {author} {\bibfnamefont {A.~R.}\ \bibnamefont
  {Ahmadi}},\ }\bibfield  {title} {\bibinfo {title} {An experimental
  investigation of blade-vortex interaction at normal incidence},\ }\href@noop
  {} {\bibfield  {journal} {\bibinfo  {journal} {Journal of Aircraft}\ }\textbf
  {\bibinfo {volume} {23}},\ \bibinfo {pages} {47} (\bibinfo {year}
  {1986})}\BibitemShut {NoStop}%
\bibitem [{\citenamefont {Kim}\ and\ \citenamefont
  {Komerath}(1995)}]{kim1995summary}%
  \BibitemOpen
  \bibfield  {author} {\bibinfo {author} {\bibfnamefont {J.}~\bibnamefont
  {Kim}}\ and\ \bibinfo {author} {\bibfnamefont {N.}~\bibnamefont {Komerath}},\
  }\bibfield  {title} {\bibinfo {title} {Summary of the interaction of a rotor
  wake with a circular cylinder},\ }\href@noop {} {\bibfield  {journal}
  {\bibinfo  {journal} {AIAA journal}\ }\textbf {\bibinfo {volume} {33}},\
  \bibinfo {pages} {470} (\bibinfo {year} {1995})}\BibitemShut {NoStop}%
\bibitem [{\citenamefont {Krishnamoorthy}\ and\ \citenamefont
  {Marshall}(1998)}]{krishnamoorthy1998three}%
  \BibitemOpen
  \bibfield  {author} {\bibinfo {author} {\bibfnamefont {S.}~\bibnamefont
  {Krishnamoorthy}}\ and\ \bibinfo {author} {\bibfnamefont {J.}~\bibnamefont
  {Marshall}},\ }\bibfield  {title} {\bibinfo {title} {Three-dimensional
  blade--vortex interaction in the strong vortex regime},\ }\href@noop {}
  {\bibfield  {journal} {\bibinfo  {journal} {Physics of Fluids}\ }\textbf
  {\bibinfo {volume} {10}},\ \bibinfo {pages} {2828} (\bibinfo {year}
  {1998})}\BibitemShut {NoStop}%
\bibitem [{\citenamefont {Gossler}\ and\ \citenamefont
  {Marshall}(2001)}]{gossler2001simulation}%
  \BibitemOpen
  \bibfield  {author} {\bibinfo {author} {\bibfnamefont {A.}~\bibnamefont
  {Gossler}}\ and\ \bibinfo {author} {\bibfnamefont {J.}~\bibnamefont
  {Marshall}},\ }\bibfield  {title} {\bibinfo {title} {Simulation of normal
  vortex--cylinder interaction in a viscous fluid},\ }\href@noop {} {\bibfield
  {journal} {\bibinfo  {journal} {Journal of Fluid Mechanics}\ }\textbf
  {\bibinfo {volume} {431}},\ \bibinfo {pages} {371} (\bibinfo {year}
  {2001})}\BibitemShut {NoStop}%
\bibitem [{\citenamefont {Felli}\ \emph {et~al.}(2009)\citenamefont {Felli},
  \citenamefont {Roberto},\ and\ \citenamefont {Guj}}]{felli2009experimental}%
  \BibitemOpen
  \bibfield  {author} {\bibinfo {author} {\bibfnamefont {M.}~\bibnamefont
  {Felli}}, \bibinfo {author} {\bibfnamefont {C.}~\bibnamefont {Roberto}},\
  and\ \bibinfo {author} {\bibfnamefont {G.}~\bibnamefont {Guj}},\ }\bibfield
  {title} {\bibinfo {title} {Experimental analysis of the flow field around a
  propeller--rudder configuration},\ }\href@noop {} {\bibfield  {journal}
  {\bibinfo  {journal} {Experiments in Fluids}\ }\textbf {\bibinfo {volume}
  {46}},\ \bibinfo {pages} {147} (\bibinfo {year} {2009})}\BibitemShut
  {NoStop}%
\bibitem [{\citenamefont {Peng}\ and\ \citenamefont
  {Gregory}(2015)}]{peng2015vortex}%
  \BibitemOpen
  \bibfield  {author} {\bibinfo {author} {\bibfnamefont {D.}~\bibnamefont
  {Peng}}\ and\ \bibinfo {author} {\bibfnamefont {J.~W.}\ \bibnamefont
  {Gregory}},\ }\bibfield  {title} {\bibinfo {title} {Vortex dynamics during
  blade-vortex interactions},\ }\href@noop {} {\bibfield  {journal} {\bibinfo
  {journal} {Physics of Fluids}\ }\textbf {\bibinfo {volume} {27}},\ \bibinfo
  {pages} {053104} (\bibinfo {year} {2015})}\BibitemShut {NoStop}%
\bibitem [{\citenamefont {Felli}(2021)}]{felli2021underlying}%
  \BibitemOpen
  \bibfield  {author} {\bibinfo {author} {\bibfnamefont {M.}~\bibnamefont
  {Felli}},\ }\bibfield  {title} {\bibinfo {title} {Underlying mechanisms of
  propeller wake interaction with a wing},\ }\href@noop {} {\bibfield
  {journal} {\bibinfo  {journal} {Journal of Fluid Mechanics}\ }\textbf
  {\bibinfo {volume} {908}},\ \bibinfo {pages} {A10} (\bibinfo {year}
  {2021})}\BibitemShut {NoStop}%
\bibitem [{\citenamefont {Marshall}\ and\ \citenamefont
  {Krishnamoorthy}(1997)}]{marshall1997instantaneous}%
  \BibitemOpen
  \bibfield  {author} {\bibinfo {author} {\bibfnamefont {J.~S.}\ \bibnamefont
  {Marshall}}\ and\ \bibinfo {author} {\bibfnamefont {S.}~\bibnamefont
  {Krishnamoorthy}},\ }\bibfield  {title} {\bibinfo {title} {On the
  instantaneous cutting of a columnar vortex with non-zero axial flow},\
  }\href@noop {} {\bibfield  {journal} {\bibinfo  {journal} {Journal of Fluid
  Mechanics}\ }\textbf {\bibinfo {volume} {351}},\ \bibinfo {pages} {41}
  (\bibinfo {year} {1997})}\BibitemShut {NoStop}%
\bibitem [{\citenamefont {Affes}\ and\ \citenamefont
  {Conlisk}(1993)}]{affes1993model_part_1}%
  \BibitemOpen
  \bibfield  {author} {\bibinfo {author} {\bibfnamefont {H.}~\bibnamefont
  {Affes}}\ and\ \bibinfo {author} {\bibfnamefont {A.}~\bibnamefont
  {Conlisk}},\ }\bibfield  {title} {\bibinfo {title} {Model for rotor tip
  vortex-airframe interaction. {{{I-T}}}heory},\ }\href@noop {} {\bibfield
  {journal} {\bibinfo  {journal} {AIAA journal}\ }\textbf {\bibinfo {volume}
  {31}},\ \bibinfo {pages} {2263} (\bibinfo {year} {1993})}\BibitemShut
  {NoStop}%
\bibitem [{\citenamefont {Moore}\ and\ \citenamefont
  {Saffman}(1972)}]{moore1972motion}%
  \BibitemOpen
  \bibfield  {author} {\bibinfo {author} {\bibfnamefont {D.~W.}\ \bibnamefont
  {Moore}}\ and\ \bibinfo {author} {\bibfnamefont {P.~G.}\ \bibnamefont
  {Saffman}},\ }\bibfield  {title} {\bibinfo {title} {The motion of a vortex
  filament with axial flow},\ }\href@noop {} {\bibfield  {journal} {\bibinfo
  {journal} {Philosophical Transactions of the Royal Society of London. Series
  A, Mathematical and Physical Sciences}\ }\textbf {\bibinfo {volume} {272}},\
  \bibinfo {pages} {403} (\bibinfo {year} {1972})}\BibitemShut {NoStop}%
\bibitem [{\citenamefont {Lundgren}\ and\ \citenamefont
  {Ashurst}(1989)}]{lundgren1989area}%
  \BibitemOpen
  \bibfield  {author} {\bibinfo {author} {\bibfnamefont {T.}~\bibnamefont
  {Lundgren}}\ and\ \bibinfo {author} {\bibfnamefont {W.}~\bibnamefont
  {Ashurst}},\ }\bibfield  {title} {\bibinfo {title} {Area-varying waves on
  curved vortex tubes with application to vortex breakdown},\ }\href@noop {}
  {\bibfield  {journal} {\bibinfo  {journal} {Journal of Fluid Mechanics}\
  }\textbf {\bibinfo {volume} {200}},\ \bibinfo {pages} {283} (\bibinfo {year}
  {1989})}\BibitemShut {NoStop}%
\bibitem [{\citenamefont {Marshall}(1991)}]{marshall1991general}%
  \BibitemOpen
  \bibfield  {author} {\bibinfo {author} {\bibfnamefont {J.}~\bibnamefont
  {Marshall}},\ }\bibfield  {title} {\bibinfo {title} {A general theory of
  curved vortices with circular cross-section and variable core area},\
  }\href@noop {} {\bibfield  {journal} {\bibinfo  {journal} {Journal of Fluid
  Mechanics}\ }\textbf {\bibinfo {volume} {229}},\ \bibinfo {pages} {311}
  (\bibinfo {year} {1991})}\BibitemShut {NoStop}%
\bibitem [{\citenamefont {Affes}\ \emph {et~al.}(1993)\citenamefont {Affes},
  \citenamefont {Conlisk}, \citenamefont {Kim},\ and\ \citenamefont
  {Komerath}}]{affes1993model_part_2}%
  \BibitemOpen
  \bibfield  {author} {\bibinfo {author} {\bibfnamefont {H.}~\bibnamefont
  {Affes}}, \bibinfo {author} {\bibfnamefont {A.}~\bibnamefont {Conlisk}},
  \bibinfo {author} {\bibfnamefont {J.}~\bibnamefont {Kim}},\ and\ \bibinfo
  {author} {\bibfnamefont {N.}~\bibnamefont {Komerath}},\ }\bibfield  {title}
  {\bibinfo {title} {Model for rotor tip vortex-airframe interaction.
  {{{II-C}}}omparison with experiment},\ }\href@noop {} {\bibfield  {journal}
  {\bibinfo  {journal} {AIAA journal}\ }\textbf {\bibinfo {volume} {31}},\
  \bibinfo {pages} {2274} (\bibinfo {year} {1993})}\BibitemShut {NoStop}%
\bibitem [{\citenamefont {Doligalski}\ and\ \citenamefont
  {Walker}(1984)}]{doligalski1984boundary}%
  \BibitemOpen
  \bibfield  {author} {\bibinfo {author} {\bibfnamefont {T.}~\bibnamefont
  {Doligalski}}\ and\ \bibinfo {author} {\bibfnamefont {J.}~\bibnamefont
  {Walker}},\ }\bibfield  {title} {\bibinfo {title} {The boundary layer induced
  by a convected two-dimensional vortex},\ }\href@noop {} {\bibfield  {journal}
  {\bibinfo  {journal} {Journal of Fluid Mechanics}\ }\textbf {\bibinfo
  {volume} {139}},\ \bibinfo {pages} {1} (\bibinfo {year} {1984})}\BibitemShut
  {NoStop}%
\bibitem [{\citenamefont {Luton}\ \emph {et~al.}(1995)\citenamefont {Luton},
  \citenamefont {Ragab},\ and\ \citenamefont
  {Telionis}}]{Luton1995InteractionOS}%
  \BibitemOpen
  \bibfield  {author} {\bibinfo {author} {\bibfnamefont {A.}~\bibnamefont
  {Luton}}, \bibinfo {author} {\bibfnamefont {S.}~\bibnamefont {Ragab}},\ and\
  \bibinfo {author} {\bibfnamefont {D.~P.}\ \bibnamefont {Telionis}},\
  }\bibfield  {title} {\bibinfo {title} {Interaction of spanwise vortices with
  a boundary layer},\ }\href@noop {} {\bibfield  {journal} {\bibinfo  {journal}
  {Physics of Fluids}\ }\textbf {\bibinfo {volume} {7}},\ \bibinfo {pages}
  {2757} (\bibinfo {year} {1995})}\BibitemShut {NoStop}%
\bibitem [{\citenamefont {Van~Dommelen}\ and\ \citenamefont
  {Cowley}(1990)}]{van1990lagrangian}%
  \BibitemOpen
  \bibfield  {author} {\bibinfo {author} {\bibfnamefont {L.~L.}\ \bibnamefont
  {Van~Dommelen}}\ and\ \bibinfo {author} {\bibfnamefont {S.~J.}\ \bibnamefont
  {Cowley}},\ }\bibfield  {title} {\bibinfo {title} {On the {{{Lagrangian}}}
  description of unsteady boundary-layer separation. {{{Part 1. G}}}eneral
  theory},\ }\href@noop {} {\bibfield  {journal} {\bibinfo  {journal} {Journal
  of Fluid Mechanics}\ }\textbf {\bibinfo {volume} {210}},\ \bibinfo {pages}
  {593} (\bibinfo {year} {1990})}\BibitemShut {NoStop}%
\bibitem [{\citenamefont {Affes}\ \emph {et~al.}(1994)\citenamefont {Affes},
  \citenamefont {Xiao},\ and\ \citenamefont {Conlisk}}]{affes1994boundary}%
  \BibitemOpen
  \bibfield  {author} {\bibinfo {author} {\bibfnamefont {H.}~\bibnamefont
  {Affes}}, \bibinfo {author} {\bibfnamefont {Z.}~\bibnamefont {Xiao}},\ and\
  \bibinfo {author} {\bibfnamefont {A.}~\bibnamefont {Conlisk}},\ }\bibfield
  {title} {\bibinfo {title} {The boundary-layer flow due to a vortex
  approaching a cylinder},\ }\href@noop {} {\bibfield  {journal} {\bibinfo
  {journal} {Journal of Fluid Mechanics}\ }\textbf {\bibinfo {volume} {275}},\
  \bibinfo {pages} {33} (\bibinfo {year} {1994})}\BibitemShut {NoStop}%
\bibitem [{\citenamefont {Saunders}\ and\ \citenamefont
  {Marshall}(2015)}]{saunders2015vorticity}%
  \BibitemOpen
  \bibfield  {author} {\bibinfo {author} {\bibfnamefont {D.~C.}\ \bibnamefont
  {Saunders}}\ and\ \bibinfo {author} {\bibfnamefont {J.~S.}\ \bibnamefont
  {Marshall}},\ }\bibfield  {title} {\bibinfo {title} {Vorticity reconnection
  during vortex cutting by a blade},\ }\href@noop {} {\bibfield  {journal}
  {\bibinfo  {journal} {Journal of Fluid Mechanics}\ }\textbf {\bibinfo
  {volume} {782}},\ \bibinfo {pages} {37} (\bibinfo {year} {2015})}\BibitemShut
  {NoStop}%
\bibitem [{\citenamefont {Leweke}\ \emph {et~al.}(2016)\citenamefont {Leweke},
  \citenamefont {Le~Dizes},\ and\ \citenamefont
  {Williamson}}]{leweke2016dynamics}%
  \BibitemOpen
  \bibfield  {author} {\bibinfo {author} {\bibfnamefont {T.}~\bibnamefont
  {Leweke}}, \bibinfo {author} {\bibfnamefont {S.}~\bibnamefont {Le~Dizes}},\
  and\ \bibinfo {author} {\bibfnamefont {C.~H.}\ \bibnamefont {Williamson}},\
  }\bibfield  {title} {\bibinfo {title} {Dynamics and instabilities of vortex
  pairs},\ }\href@noop {} {\bibfield  {journal} {\bibinfo  {journal} {Annual
  Review of Fluid Mechanics}\ }\textbf {\bibinfo {volume} {48}},\ \bibinfo
  {pages} {507} (\bibinfo {year} {2016})}\BibitemShut {NoStop}%
\bibitem [{\citenamefont {Van Der~Poel}\ \emph {et~al.}(2015)\citenamefont {Van
  Der~Poel}, \citenamefont {Ostilla-M{\'o}nico}, \citenamefont {Donners},\ and\
  \citenamefont {Verzicco}}]{van2015pencil}%
  \BibitemOpen
  \bibfield  {author} {\bibinfo {author} {\bibfnamefont {E.~P.}\ \bibnamefont
  {Van Der~Poel}}, \bibinfo {author} {\bibfnamefont {R.}~\bibnamefont
  {Ostilla-M{\'o}nico}}, \bibinfo {author} {\bibfnamefont {J.}~\bibnamefont
  {Donners}},\ and\ \bibinfo {author} {\bibfnamefont {R.}~\bibnamefont
  {Verzicco}},\ }\bibfield  {title} {\bibinfo {title} {A pencil distributed
  finite difference code for strongly turbulent wall-bounded flows},\
  }\href@noop {} {\bibfield  {journal} {\bibinfo  {journal} {Computers \&
  Fluids}\ }\textbf {\bibinfo {volume} {116}},\ \bibinfo {pages} {10} (\bibinfo
  {year} {2015})}\BibitemShut {NoStop}%
\bibitem [{\citenamefont {Spandan}\ \emph {et~al.}(2017)\citenamefont
  {Spandan}, \citenamefont {Meschini}, \citenamefont {Ostilla-M{\'o}nico},
  \citenamefont {Lohse}, \citenamefont {Querzoli}, \citenamefont {de~Tullio},\
  and\ \citenamefont {Verzicco}}]{spandan2017parallel}%
  \BibitemOpen
  \bibfield  {author} {\bibinfo {author} {\bibfnamefont {V.}~\bibnamefont
  {Spandan}}, \bibinfo {author} {\bibfnamefont {V.}~\bibnamefont {Meschini}},
  \bibinfo {author} {\bibfnamefont {R.}~\bibnamefont {Ostilla-M{\'o}nico}},
  \bibinfo {author} {\bibfnamefont {D.}~\bibnamefont {Lohse}}, \bibinfo
  {author} {\bibfnamefont {G.}~\bibnamefont {Querzoli}}, \bibinfo {author}
  {\bibfnamefont {M.~D.}\ \bibnamefont {de~Tullio}},\ and\ \bibinfo {author}
  {\bibfnamefont {R.}~\bibnamefont {Verzicco}},\ }\bibfield  {title} {\bibinfo
  {title} {A parallel interaction potential approach coupled with the immersed
  boundary method for fully resolved simulations of deformable interfaces and
  membranes},\ }\href@noop {} {\bibfield  {journal} {\bibinfo  {journal}
  {Journal of Computational Physics}\ }\textbf {\bibinfo {volume} {348}},\
  \bibinfo {pages} {567} (\bibinfo {year} {2017})}\BibitemShut {NoStop}%
\bibitem [{\citenamefont {Spandan}\ \emph {et~al.}(2018)\citenamefont
  {Spandan}, \citenamefont {Verzicco},\ and\ \citenamefont
  {Lohse}}]{spandan2018physical}%
  \BibitemOpen
  \bibfield  {author} {\bibinfo {author} {\bibfnamefont {V.}~\bibnamefont
  {Spandan}}, \bibinfo {author} {\bibfnamefont {R.}~\bibnamefont {Verzicco}},\
  and\ \bibinfo {author} {\bibfnamefont {D.}~\bibnamefont {Lohse}},\ }\bibfield
   {title} {\bibinfo {title} {Physical mechanisms governing drag reduction in
  turbulent taylor--couette flow with finite-size deformable bubbles},\
  }\href@noop {} {\bibfield  {journal} {\bibinfo  {journal} {Journal of fluid
  mechanics}\ }\textbf {\bibinfo {volume} {849}},\ \bibinfo {pages} {R3}
  (\bibinfo {year} {2018})}\BibitemShut {NoStop}%
\bibitem [{\citenamefont {Zhu}\ \emph {et~al.}(2018)\citenamefont {Zhu},
  \citenamefont {Verschoof}, \citenamefont {Bakhuis}, \citenamefont {Huisman},
  \citenamefont {Verzicco}, \citenamefont {Sun},\ and\ \citenamefont
  {Lohse}}]{zhu2018wall}%
  \BibitemOpen
  \bibfield  {author} {\bibinfo {author} {\bibfnamefont {X.}~\bibnamefont
  {Zhu}}, \bibinfo {author} {\bibfnamefont {R.~A.}\ \bibnamefont {Verschoof}},
  \bibinfo {author} {\bibfnamefont {D.}~\bibnamefont {Bakhuis}}, \bibinfo
  {author} {\bibfnamefont {S.~G.}\ \bibnamefont {Huisman}}, \bibinfo {author}
  {\bibfnamefont {R.}~\bibnamefont {Verzicco}}, \bibinfo {author}
  {\bibfnamefont {C.}~\bibnamefont {Sun}},\ and\ \bibinfo {author}
  {\bibfnamefont {D.}~\bibnamefont {Lohse}},\ }\bibfield  {title} {\bibinfo
  {title} {Wall roughness induces asymptotic ultimate turbulence},\ }\href@noop
  {} {\bibfield  {journal} {\bibinfo  {journal} {Nature physics}\ }\textbf
  {\bibinfo {volume} {14}},\ \bibinfo {pages} {417} (\bibinfo {year}
  {2018})}\BibitemShut {NoStop}%
\bibitem [{\citenamefont {Berghout}\ \emph {et~al.}(2019)\citenamefont
  {Berghout}, \citenamefont {Zhu}, \citenamefont {Chung}, \citenamefont
  {Verzicco}, \citenamefont {Stevens},\ and\ \citenamefont
  {Lohse}}]{berghout2019direct}%
  \BibitemOpen
  \bibfield  {author} {\bibinfo {author} {\bibfnamefont {P.}~\bibnamefont
  {Berghout}}, \bibinfo {author} {\bibfnamefont {X.}~\bibnamefont {Zhu}},
  \bibinfo {author} {\bibfnamefont {D.}~\bibnamefont {Chung}}, \bibinfo
  {author} {\bibfnamefont {R.}~\bibnamefont {Verzicco}}, \bibinfo {author}
  {\bibfnamefont {R.~J.}\ \bibnamefont {Stevens}},\ and\ \bibinfo {author}
  {\bibfnamefont {D.}~\bibnamefont {Lohse}},\ }\bibfield  {title} {\bibinfo
  {title} {Direct numerical simulations of taylor--couette turbulence: the
  effects of sand grain roughness},\ }\href@noop {} {\bibfield  {journal}
  {\bibinfo  {journal} {Journal of fluid mechanics}\ }\textbf {\bibinfo
  {volume} {873}},\ \bibinfo {pages} {260} (\bibinfo {year}
  {2019})}\BibitemShut {NoStop}%
\bibitem [{\citenamefont {Hunt}\ \emph {et~al.}(1988)\citenamefont {Hunt},
  \citenamefont {Wray},\ and\ \citenamefont {Moin}}]{hunt1988eddies}%
  \BibitemOpen
  \bibfield  {author} {\bibinfo {author} {\bibfnamefont {J.~C.}\ \bibnamefont
  {Hunt}}, \bibinfo {author} {\bibfnamefont {A.~A.}\ \bibnamefont {Wray}},\
  and\ \bibinfo {author} {\bibfnamefont {P.}~\bibnamefont {Moin}},\ }\bibfield
  {title} {\bibinfo {title} {Eddies, streams, and convergence zones in
  turbulent flows},\ }\href@noop {} {\bibfield  {journal} {\bibinfo  {journal}
  {Studying turbulence using numerical simulation databases, 2. Proceedings of
  the 1988 summer program}\ } (\bibinfo {year} {1988})}\BibitemShut {NoStop}%
\bibitem [{\citenamefont {Meneveau}(2011)}]{meneveau2011lagrangian}%
  \BibitemOpen
  \bibfield  {author} {\bibinfo {author} {\bibfnamefont {C.}~\bibnamefont
  {Meneveau}},\ }\bibfield  {title} {\bibinfo {title} {Lagrangian dynamics and
  models of the velocity gradient tensor in turbulent flows},\ }\href@noop {}
  {\bibfield  {journal} {\bibinfo  {journal} {Annual Review of Fluid
  Mechanics}\ }\textbf {\bibinfo {volume} {43}},\ \bibinfo {pages} {219}
  (\bibinfo {year} {2011})}\BibitemShut {NoStop}%
\bibitem [{\citenamefont {Pumir}(1994)}]{pumir1994numerical}%
  \BibitemOpen
  \bibfield  {author} {\bibinfo {author} {\bibfnamefont {A.}~\bibnamefont
  {Pumir}},\ }\bibfield  {title} {\bibinfo {title} {A numerical study of
  pressure fluctuations in three-dimensional, incompressible, homogeneous,
  isotropic turbulence},\ }\href@noop {} {\bibfield  {journal} {\bibinfo
  {journal} {Physics of Fluids}\ }\textbf {\bibinfo {volume} {6}},\ \bibinfo
  {pages} {2071} (\bibinfo {year} {1994})}\BibitemShut {NoStop}%
\bibitem [{\citenamefont {Hussain}\ and\ \citenamefont
  {Duraisamy}(2011)}]{hussain2011mechanics}%
  \BibitemOpen
  \bibfield  {author} {\bibinfo {author} {\bibfnamefont {F.}~\bibnamefont
  {Hussain}}\ and\ \bibinfo {author} {\bibfnamefont {K.}~\bibnamefont
  {Duraisamy}},\ }\bibfield  {title} {\bibinfo {title} {Mechanics of viscous
  vortex reconnection},\ }\href@noop {} {\bibfield  {journal} {\bibinfo
  {journal} {Physics of Fluids}\ }\textbf {\bibinfo {volume} {23}} (\bibinfo
  {year} {2011})}\BibitemShut {NoStop}%
\end{thebibliography}%

\end{document}